# Scientific-Intention Driven Embodied Intelligent Solar Telescope: Conceptual Design

Lin Jiaben[1]*, Tong Liyue[1], Wang Hui[1,2], Shao Mingfu[1,2], Yang Chen[1,2]

[1]State Key Laboratory of Solar Activity and Space Weather, National Astronomical Observatories, Chinese Academy of Sciences, Beijing 100101, China

[2]University of Chinese Academy of Sciences, Beijing 100049, China

## Abstract

Artificial Intelligence (AI) is profoundly transforming the paradigms of scientific research. Cutting-edge technologies such as Large Language Models (LLMs) and embodied intelligence are continuously pushing the boundaries of scientific instrumentation. Against this backdrop, this paper proposes a novel conceptual system: the Scientific-Intention Driven Embodied Intelligent Solar Telescope (SIDEST). The system is designed with three core layers to achieve three types of intelligent scientific research closed loops. First, the Scientific Intent Research and Demonstration Layer parses the research objectives and intents of scientists (e.g., solar physicists) through natural language interaction, achieving a closed loop for the generation and optimization of executable observation plans aligned with scientific intent via in-depth research. Subsequently, the Observation Realization Layer schedules embodied intelligent solar telescopes to implement a closed loop for the execution of scientific observation plans. Finally, the Evaluation and Evolution Layer coordinates intelligent agents for data processing and scientific analysis to analyze observation data, generate research reports, and iteratively optimize observation strategies and model methods based on results, thereby realizing a self-evolving closed loop for the entire system. During the research process, we constructed a minimal prototype system based on a precision temperature control device for solar telescope birefringent filters to validate the core principles of SIDEST. This prototype successfully implemented all key steps of intention-driven automated research, demonstrating the feasibility of the technical pathways for the three types of intelligent research closed loops. SIDEST redefines telescopes through cutting-edge AI methods. By integrating AI's reasoning and planning capabilities with scientists' creative thinking, it establishes a new "human-machine collaborative" approach to astronomical research, exploring possible development directions and feasible technical pathways for next-generation astronomical telescopes, intelligent instruments, and new paradigms of astronomical research.

## 1 Introduction

The Sun is the central body of the solar system, governing its evolutionary processes and providing indispensable energy for life on Earth. It is not only a key subject of astrophysical research but also a storm source requiring continuous monitoring for space weather forecasting. With the increasing sophistication and spaceward expansion of modern high-technology systems, solar activity—especially solar storms—has an increasingly pronounced potential for disruptive impacts on human technological society. To address these possible effects and challenges, countries worldwide have deployed extensive space-based and ground-based solar monitoring instruments, such as the Solar Dynamics Observatory (SDO) [1] and the Advanced Space-based Solar Observatory (ASO-S, also known as Kuafu-1) [2]. These observational facilities have propelled solar observations into the era of terabyte-scale daily data volumes, providing researchers with unprecedented research opportunities. Although both observational technology and data processing and analysis techniques have undergone transformative advances, the current astronomical research methodology—including solar physics research—essentially still follows the paradigm established since the Galilean era: astronomers propose hypotheses, design observation plans [3–4], and drive scientific discovery through manual data analysis. This research approach is heavily dependent on individual experience and intuition, and its efficiency and scientific output capacity are increasingly inadequate for the demands of massive data volumes and complex scientific problems. Meanwhile, related studies indicate that under such research paradigms, the rate of progress in most scientific disciplines declines by approximately half every 13 years [5]. Thus, whether viewed from the perspective of solar physics and space weather research needs, or from the developmental trends of research methodology itself, there is an urgent need for new paradigms to substantially improve research efficiency and capability.

In recent years, artificial intelligence technologies represented by Large Language Models (LLMs) have demonstrated remarkable capabilities in knowledge understanding, task planning, and logical reasoning [6–10], continuously integrating with various scientific disciplines and becoming essential tools in scientific research. Building upon general-purpose LLMs, researchers in chemistry, biomedicine, and many other fields have constructed their own domain-specific large models and agent systems. These systems possess the capability to autonomously complete the full pipeline of "hypothesis generation–experimental verification–result analysis," thereby constituting genuinely automated scientific research systems [11–13, 27–29]. However, these systems are primarily designed to address well-defined types of scientific problems. Can a system be built that understands research intentions expressed by researchers in natural language and autonomously conducts broad-domain, multi-directional investigations?

In practice, the development of LLMs has provided an efficient technical pathway for intent recognition tasks—through fine-tuning foundation models, designing specialized prompts, and employing retrieval-augmented generation (RAG) methods, high-accuracy intent recognition can be achieved. Concurrently, embodied intelligence (Embodied AI), driven by LLMs as its core, is also undergoing rapid development. When this concept is applied to robotics and experimental science, it gives rise to intelligent instrumentation systems capable of autonomously conducting experiments. Such systems are no longer confined to passively executing pre-defined instructions; instead, they can achieve efficient experimentation and research through interactive understanding of environments and experimental conditions [14]. For example, in the fields of biochemistry and novel materials research, by integrating robotic experiment platforms, multi-modal sensing modules, adaptive control algorithms, and the cognitive reasoning capabilities of LLMs, continuously operating autonomous research systems have been constructed, significantly enhancing the efficiency of novel material screening and chemical reaction pathway optimization [15]. In astronomy, the StarWhisper Telescope (SWT) system, based on an LLM-driven intelligent agent framework, has achieved automated management of astronomical observations. Applied to the Nearby Galaxy Supernova Survey (NGSS) project, it integrates observation plan generation, telescope control, data processing, and real-time feedback functions, lowering operational barriers through natural language interaction and effectively reducing the workload of astronomers [24]. Such systems represent an important evolutionary direction for scientific research paradigms—namely, AI-driven autonomous scientific research—and also demonstrate the practical potential of multi-agent collaboration in experimental environments.

Based on the above analysis of the technological development landscape, we propose the "Scientific-Intention Driven Embodied Intelligent Solar Telescope" (SIDEST), an intelligent observation system oriented toward next-generation solar physics research. Its core concept is to endow the telescope system with the ability to understand scientific intent, enabling it to automatically transform research questions expressed by scientists in natural language (e.g., "investigate the triggering mechanism of solar flares," "analyze the relationship between active region magnetic field evolution and filament eruption") into executable observation tasks, and to realize a full-process research closed loop encompassing task planning, observation execution, data acquisition, and result analysis. The SIDEST system will be constructed based on the existing telescope facilities at the Huairou Solar Observing Station of the National Astronomical Observatories, including a 35 cm solar magnetic field telescope [16], a full-disk vector magnetograph, and a full-disk chromospheric (Hα) telescope [17]. Each telescope will undergo embodied intelligence retrofitting and upgrading. Through deep integration of LLMs, multi-modal perception, adaptive control, in-depth research, and other AI technologies, SIDEST will form an intelligent observational research system with autonomous observation, dynamic decision-making, and continuous evolution capabilities. The system will not only be able to recognize and respond to scientific objectives proposed by astronomers, but also automatically optimize observation strategies during observations based on real-time environmental conditions, telescope operational status, and feedback from observation targets, enabling intelligent tracking and in-depth investigation of solar activity phenomena, and realizing the paradigm shift of astronomical observational research from "human-driven" to "human-machine collaborative" intelligence.

The remainder of this paper is organized as follows: Section 2 provides an in-depth discussion of the fundamental content of the SIDEST concept; Section 3 introduces the basic design framework of SIDEST; Section 4 presents the prototype system experiments and results; and finally, Section 5 discusses and envisions future development directions for this concept.

## 2 The Scientific-Intention Driven Embodied Intelligent Solar Telescope Concept

The telescope is an irreplaceable instrument in humanity's quest to explore the universe, extending our perceptual and cognitive capabilities. For a long time, telescopes have functioned as passive execution tools, strictly following procedures preset by astronomers and engineers to perform actions and produce scientific data. In recent years, with the rapid development of automatic control, robotics, and related technologies—especially breakthroughs in artificial intelligence represented by LLMs and intelligent agents—astronomical telescopes, like other scientific instruments, are undergoing a paradigm shift toward embodied intelligence. Under this new paradigm, telescopes (and scientific instruments) will no longer be devices that passively execute instructions, but rather intelligent machines capable of actively analyzing tasks and environments, and achieving superior execution schemes through physical interaction [18]. In fact, as early as 1950, Turing proposed the concept of embodied intelligence, emphasizing that intelligence must achieve autonomous learning and evolution through dynamic interaction between embodiment and environment, with its core lying in the deep integration of perception, action, and cognition. Applying this philosophy to astronomical observation equipment gives rise to the concept of the embodied telescope. Such telescopes possess dual perceptual capabilities—internal and external: internally, they can monitor in real time the full chain of instrumental state information including pointing errors, temperature variations, and instrumental parameters of the telescope itself; externally, they can accurately perceive meteorological conditions, cloud cover distribution, and other external environmental factors that may affect observations. Building on this foundation, through deep integration of AI technologies, the leap from embodiment to intelligence is achieved, thereby giving birth to the embodied intelligent telescope. It is no longer merely a device that automatically executes preset plans, but an advanced intelligent system capable of autonomously making decisions, planning, and executing observation tasks based on real-time environmental conditions, instrumental status, and scientific mission requirements.

Building upon the theoretical and practical foundation of embodied intelligent telescopes, this study further proposes the conceptual system of the "Scientific-Intention Driven Embodied Intelligent Solar Telescope." This system, grounded in cutting-edge AI technologies, aims to achieve systematic enhancement of solar telescope observational capabilities. Its core innovation lies in transcending the limitation of traditional embodied intelligence, which primarily focuses on physical environment interaction, by introducing the recognition, understanding, and logical demonstration of scientists' research intentions, thereby enabling higher-level interaction and co-evolution between observation execution and scientific research intent—transforming the observing instrument from a passive data acquisition tool into an active scientific discovery partner.

The system design architecture of SIDEST can be abstracted into three core layers, constituting a complete perception–decision–action–evaluation closed loop:

Scientific Intent Research and Demonstration Layer: This layer corresponds to the process from scientific intent understanding to the generation and demonstration of verifiable scientific hypotheses. As the entry point of the SIDEST system, it leverages the deep comprehension capabilities of LLMs for natural language observation requests to construct an intelligent system capable of understanding scientists' research intentions. Through natural language processing and knowledge graph technologies, it parses scientific questions [19, 20], generates executable research hypotheses, and transforms them into observation targets and planning schemes, thereby achieving an optimization closed loop for observation planning. In traditional scientific workflows, translating scientific intents into executable plans typically requires scientists and engineering personnel to invest substantial time in demonstrating the feasibility of a series of possible hypothesis space solutions that could support the intent, and validating their effectiveness through simulations before specific observation plans can be formulated. Astronomy, as a highly interdisciplinary field, often requires multi-domain experts to collaboratively demonstrate potential hypotheses—a process that is complex and time-consuming. SIDEST, by harnessing the powerful reasoning and planning capabilities of LLMs, can significantly accelerate the transition from research intent to the screening of high-value hypotheses.

Observation Task Realization Layer: This layer corresponds to tool invocation, driving, and task realization within observation missions. It receives specific observation instructions from the layer above, dynamically schedules and controls the telescope's hardware systems, and autonomously completes observation tasks under complex and variable environmental constraints. This layer relies on the reasoning, planning, and tool-use capabilities of LLMs, combined with the telescope embodied intelligent system control framework (which encapsulates hardware components such as the dome, pointing system, imaging devices, and related software tools as a programmable skill library), forming a complete closed loop from semantic parsing to task orchestration. The system can dynamically adjust observation priorities based on observation mission objectives, real-time weather conditions, and equipment availability, and generate compliant instruction packages for transmission to the telescope control system. Within the SIDEST technical architecture, a collaborative scheduling interface connects with the available telescope network, enabling intelligent coordination of cross-site, multi-wavelength joint observations. This collaborative mechanism is particularly important for the observational analysis of phenomena such as solar eruptions, as their study typically relies on high-resolution, continuous observational data from multiple instruments and wavelengths to capture the complete physical evolution process. SIDEST can dynamically allocate observation tasks and schedule instruments based on scientific objectives and real-time environmental conditions, optimizing equipment utilization efficiency and thereby achieving observations with higher spatial and temporal resolution.

Evaluation and Evolution Layer: This layer corresponds to the feedback closed loop of scientific research and autonomous system iteration. It analyzes the collected data, assesses whether it satisfies the hypotheses and requirements proposed by the Scientific Intent Research and Demonstration Layer, and feeds evaluation results (such as data quality, task completion degree, and newly discovered clues) back to the system. Based on this feedback, the system can automatically optimize observation strategies, adjust scientific hypotheses, and even propose new exploratory directions, thereby achieving continuous learning and autonomous evolution of the entire system. The goal of SIDEST is to construct a solar telescope system with autonomous scientific discovery capabilities. Through methods such as online reinforcement learning, it performs scientific analysis on observational data, compares it with scientific intent, and achieves efficient scientific exploration and discovery through iterative optimization. For example, in flare monitoring tasks, the system feeds observational results back to the prediction model, generates scientific discoveries through comparative analysis, thereby enhancing subsequent prediction accuracy. Related discoveries are submitted to the research team in standardized report format while simultaneously updating the system knowledge base, forming a closed loop from scientific instruments to data to new knowledge and discoveries.

## 3 System Architecture and Key Technologies

SIDEST is designed with a three-tier framework structure to realize three types of intelligent scientific research closed loops. Among these, the Intent Research and Demonstration Layer serves as the interface for system–external interaction, responsible for receiving and parsing high-level research objectives or scientific intent and progressively decomposing them into demonstrable hypotheses. The Observation Realization Layer undertakes the execution function of materializing hypotheses into specific observation tasks, ensuring high-quality scientific data acquisition and standardized output. The Evaluation and Evolution Layer, as the core evaluation unit of the system, is responsible for coordinating scientific intent with observational data, constructing a closed-loop mechanism from prediction and judgment to result feedback, thereby driving the continuous iteration of the research process and the autonomous evolution of system capabilities. The system architecture is illustrated in Fig. 1.

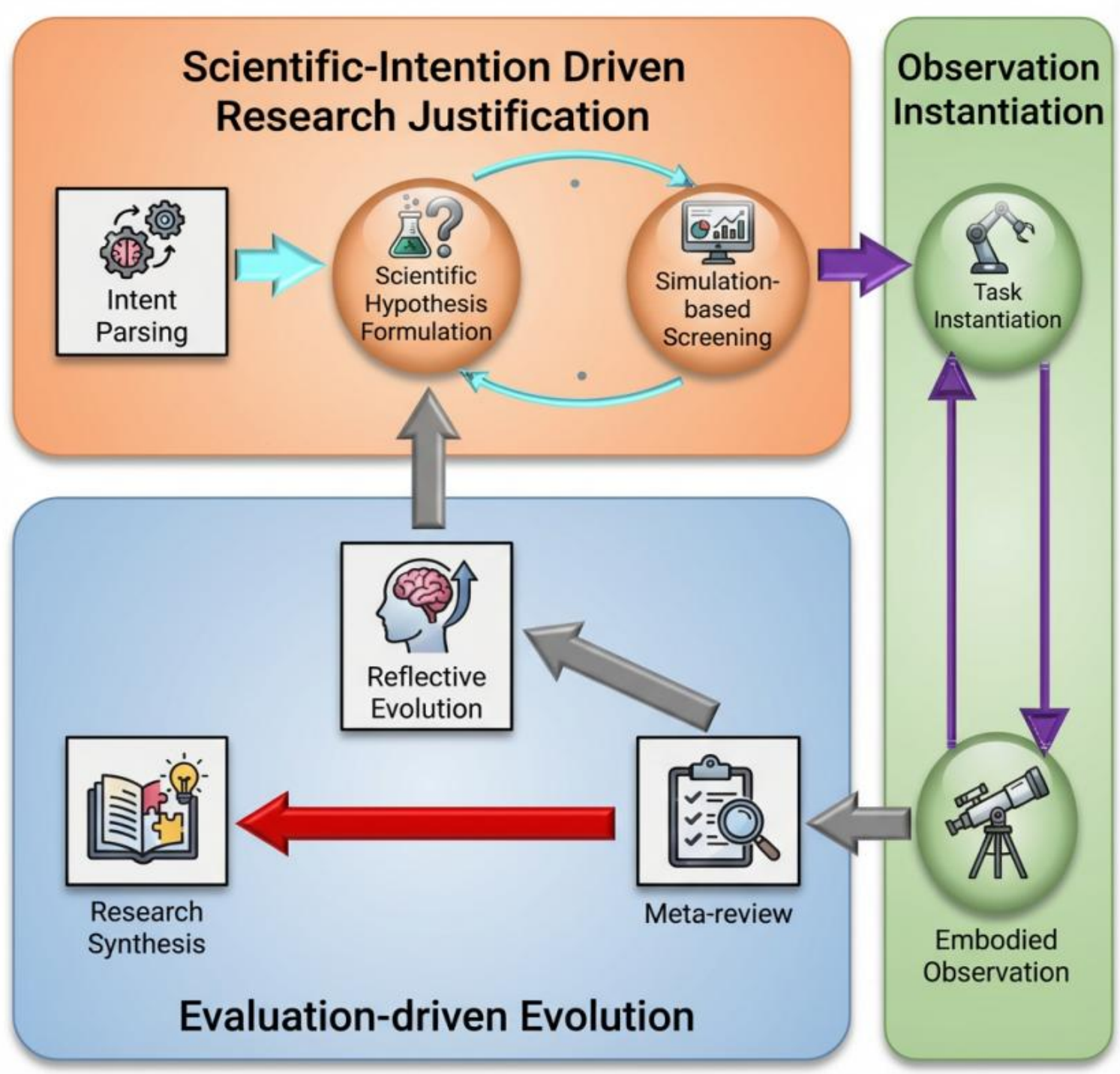

*Fig. 1 SIDEST System Architecture Diagram*

### 3.1 Overall System Framework Design

Through the three-layer architecture design, this system realizes triple closed loops spanning scientific intent parsing and optimization, embodied intelligent observation, and autonomous scientific research. The core functions and implementation pathways of each layer are described below.

#### 3.1.1 LLM-Based Scientific Intent Recognition and Research Demonstration

The concept of intent-driven systems [21] originally emerged from the field of network communications, where the goal was to automatically translate business requirements expressed in natural language by users into network configuration policies, thereby achieving automated network operation and maintenance. Subsequently, this concept gradually expanded into the field of artificial intelligence, emphasizing that systems should possess deep understanding of human goals and the ability to autonomously plan action pathways, thus becoming a core paradigm for achieving highly autonomous systems. In 2025, Academician Zheng Nanning further proposed the theoretical framework of "The Leap of Intelligence: From Model-Driven to Intent-Driven" at the WAIC conference [31]. Building upon the above research, this paper further proposes the concept of scientific-intention-driven systems and extends it to the domain of solar telescope development, resulting in the SIDEST conceptual system.

In the design of the SIDEST system, a scientific intent interface layer centered on foundation models is first constructed, supporting scientists to describe scientific objectives through natural language, graphical interfaces, or domain-specific languages—for example: "Based on parameters such as active region images, magnetic field strength, and neutral line length, predict flare eruption time and intensity [32]." Second, an intent translation and decomposition engine layer is designed. This layer is jointly constituted by LLMs, intelligent agents, and an astronomical knowledge graph. Through multi-agent-driven generation of numerous demonstrable hypotheses, combined with various pre-configured tools for simulation, feasible hypotheses are demonstrated and output to the driving interface of the next layer.

#### 3.1.2 From Observation Plan Design to Scheduling Execution: The Embodied Intelligent Telescope

The concept of embodied intelligence has been extensively researched and applied in the field of robotics. Integrating this technology into traditional astronomical observation equipment enables the construction of an embodied intelligent observation system with "perception–understanding–decision–action" closed-loop capability. This system can receive observation plans and parse them into low-level control instructions such as telescope pointing, tracking, focusing, observation terminal switching, and parameter configuration. Simultaneously, through sensors distributed across various telescope components (such as guiding sensors, position sensors, and temperature sensors), it perceives instrument status in real time and forms feedback, ensuring the precision and reliability of the execution process, thereby achieving physical closed-loop control at the hardware level.

SIDEST further enhances perception, understanding, and planning capabilities on the foundation of the embodied intelligent telescope. After obtaining demonstrable hypotheses, the system first analyzes the relevant conditions required to realize the hypotheses, including weather conditions, equipment capabilities, and whether currently available instruments and networks meet the requirements. It then proposes observationally verifiable hypotheses and materializes them into required equipment parameters and possible execution steps. Subsequently, it designs experimental schemes adapted to different hypotheses, dynamically loads and adjusts the observation task list, schedules observation plans with different priorities and conditions to their optimal states, and ultimately achieves closed-loop execution of observation tasks.

#### 3.1.3 Evolutionary Autonomous Scientific Research

Autonomous scientific research refers to a research paradigm that utilizes AI technologies (particularly LLMs and robotic systems) to achieve full-process automation of scientific research, with its core being the autonomous completion by AI agents of the complete cycle of hypothesis generation and simulation demonstration, experimental design, observation scheduling and execution, data analysis and evaluation, experimental iteration, and manuscript writing—substantially reducing human intervention. This concept can be traced back to early explorations such as the "robot scientist" and "autonomous scientific discovery" [22], with the goal of breaking through the bottlenecks of human cognition and significantly accelerating the pace of scientific discovery.

Within the SIDEST system, evolutionary autonomous scientific research is realized through AI scientist / solar physicist agents. For example, a flare prediction model or agent can predict whether a flare will erupt based on real-time magnetic field data and schedule embodied intelligent telescopes to conduct targeted observations. The observational results serve as feedback signals to optimize the prediction model and methodology—if an eruption occurs, the existing model is reinforced; if not, it triggers model reconstruction or even LLM retraining, thereby continuously improving prediction accuracy. This process fully embodies the evolutionary scientific exploration closed loop of "prediction–action–observation–learning."

Through the integration of the three closed-loop capabilities—scientific intent understanding, observation realization, and evolutionary autonomous scientific research—SIDEST constructs an intelligent solar observation system capable of conducting collaborative solar physics research. The system can carry out solar magnetic field observations, activity eruption prediction, and chromospheric eruption monitoring, forming a closed-loop logic of observation–prediction–monitoring–feedback, and realizing intent-oriented evolutionary autonomous scientific research through online reinforcement learning.

### 3.2 Key Technologies and Modules: Instruments and Data, Models, Systems

The realization of SIDEST requires key technological support across three dimensions: instrumentation and data, modeling and algorithms, and system architecture. At the instrument and data level, it is necessary to construct long-term, high-quality scientific datasets that undergo rigorous standardized preprocessing to ensure the reliability of model construction. Simultaneously, the system should possess the capability to continuously acquire new data and generate AI-ready processing results to support continuous model evolution. At the model level, deep integration of data-driven and domain-knowledge-embedded approaches is required, ensuring that models or agents both deeply understand domain knowledge and adapt to the characteristics of scientific data. Model training and agent configuration must possess flexibility to effectively integrate and leverage the advanced technologies of open-source foundation models. At the system level, a multi-agent collaborative closed-loop framework must be constructed, realizing full-process automation from scientific hypothesis generation, observation scheme design, autonomous process execution, to result analysis and iterative optimization. Furthermore, each link of the system should feature convenient human-machine interaction interfaces, ensuring that scientists can participate in and supervise the scientific research process in a "human-in-the-loop" manner.

#### 3.2.1 Key Instruments and Data Foundation

The SIDEST research builds upon the existing telescopes and data at the Huairou Observing Station, utilizing historical data to construct the system's initial algorithms and models, which then autonomously evolve during actual observations to enhance the performance metrics of system algorithms and models.

1) Vector Magnetograph and Long-Term Solar Activity Data. The 35 cm Solar Magnetic Field Telescope (completed in 1984) and the Solar Optical and Magnetic Activity Monitoring Telescope (operational since 2006) at the Huairou Observing Station have long maintained stable and reliable observations, accumulating high-precision photospheric and chromospheric vector magnetic field data spanning multiple solar cycles. Currently, we are systematically cross-calibrating these ground-based observational data with data acquired by space-borne instruments such as the Helioseismic and Magnetic Imager (HMI) onboard the Solar Dynamics Observatory (SDO) and the Full-disk MagnetoGraph (FMG) onboard the Advanced Space-based Solar Observatory (ASO-S). Through joint processing of magnetic field observations from different sources, an ultra-long-term solar activity dataset spanning several decades will be constructed. These instrumental datasets can not only provide high-quality training samples for solar physics models, but their real-time observation data streams can also serve as inputs for online reinforcement learning, driving models to dynamically respond to transient changes in solar activity and providing critical support for solar activity forecasting and space weather alerts.

2) Chromospheric Telescope and Eruption Activity Feature Data. The full-disk chromospheric (Hα) telescope at the Huairou Solar Observing Station, operational since 2006 as an essential component of the Solar Magnetism and Activity Telescope (SMAT), has been continuously providing high-quality solar chromospheric observation data. Working in coordination with the full-disk vector magnetograph installed on the same platform, this telescope acquires full-disk Hα monochromatic images at the 6562.8 Å band, enabling simultaneous observation of photospheric vector magnetic fields and chromospheric activity phenomena [23]. Its observational data are of significant value for the identification and classification of key processes such as solar flares, filament eruptions, and prominence activity. Currently, based on the long-term observations accumulated by this telescope, the SIDEST research team is constructing a specialized annotated dataset for Hα flare activity through cross-validation and labeling with GOES satellite X-ray data. This dataset will significantly enhance the reliability of solar eruption event identification and provide feedback signals for solar activity monitoring systems. Building on this, the dataset can be further integrated into the SIDEST system to support its closed-loop control workflow of observation, prediction, monitoring, and feedback optimization.

3) Domain Knowledge Base of Solar Physics Literature and Textbooks. To effectively control the "hallucination" phenomenon of LLMs and enhance their knowledge consistency within specialized domains, the SIDEST research team has systematically organized and integrated hundreds of classical textbooks and thousands of authoritative papers in the field of solar physics, constructing a structured domain knowledge base. On this foundation, a Retrieval-Augmented Generation (RAG) technical framework has been introduced, enabling models to retrieve relevant knowledge from the knowledge base in real time as grounding evidence before generating responses, thereby enhancing the scientific accuracy and reliability of outputs. In practice with the "Jinwu" series of models, this method has not only effectively improved the professional credibility of question-answering tasks but has also mitigated the issue of temporal staleness in domain knowledge caused by training data update lag. Moving forward, SIDEST research will continuously expand the scale of the knowledge base and introduce multi-modal annotations (such as causal chains, timelines, and other structured semantic information) to enhance the model's logical reasoning capability and adaptability to complex scenarios, thereby supporting demanding scientific reasoning tasks.

4) Embodied Intelligent Telescopes. The embodied intelligent telescope is a new-generation astronomical observation system that integrates multi-modal perception, autonomous decision-making, and closed-loop execution capabilities [30]. Within the SIDEST system, the planned solar telescopes have already preliminarily acquired embodied intelligence capabilities. Among these, external environmental perception is jointly provided by a compact meteorological station and an all-sky camera installed at the observing station, enabling decisions on whether to initiate or cease observations based on external meteorological conditions. The system can predict observation conditions based on meteorological station data and current all-sky cloud cover analysis, making advance observation plans. Meanwhile, the telescope's internal operational status is fed back in real time by sensors widely distributed across various telescope subsystems—such as telescope pointing position, filter temperature, filter position, system operating voltage and current status, etc. On the one hand, this allows assessment of the system's current observational capability, such as whether telescope azimuth, elevation, and the status of various instrument components meet observational requirements; on the other hand, it enables evaluation of energy storage and safety conditions under different observation modes, ensuring that planned observation tasks can be completed as scheduled.

### 3.2.2 Model and Agent Construction

The SIDEST system operates through a group of specialized models and agents coordinated by a supervisory agent, simulating the manner in which scientists conduct research. These agents, built around three major links—intent understanding and planning, observation execution, and feedback iteration—form a closed-loop scientific research paradigm. Their specific functional design is as follows:

1) Intent Parsing Agent: Perception and Parsing of Scientific Intent. This agent serves as the system input interface, responsible for parsing research objectives proposed by scientists in textual or spoken form (e.g., "study the triggering mechanism of solar flares") through natural language processing technology. It can identify key parameters, implicit constraints, and priorities within instructions, transforming them into structured, actionable research task descriptions. Its core capabilities include multi-turn dialogue context maintenance, domain terminology disambiguation, and intent confidence assessment, providing precise input foundations and background investigation for subsequent task decomposition.

2) Scientific Hypothesis Agent: Hypothesis Innovation and Cross-Domain Integration. By integrating web search, literature databases, and domain knowledge bases, this agent can generate preliminary research hypotheses and experimental directions, and enrich hypothesis content across domains. For instance, for a solar flare prediction task, it can synthesize historical eruption data, magnetic field evolution models, and the latest research findings to propose innovative ideas such as "a joint early-warning indicator based on magnetic field shear gradient and ultraviolet brightness variation."

3) Simulation Screening Agent: Scientific Demonstration and Screening of High-Value Hypotheses. This agent can simulate the scientific debate process, optimizing the plausibility and novelty of hypotheses through multiple iterative rounds, and then demonstrating the feasibility of these hypotheses through simulation—assessing required resources, time, etc.—to comprehensively evaluate and identify the high-value hypotheses closest to realizing the current intent, which then serve as input for subsequent empirical testing.

4) Task Embodiment Agent: Design from Hypothesis to Observation Task. Responsible for transforming screened hypotheses into executable observation experiments. This agent designs optimal observation sequences and parameters based on astronomical observation constraints (such as equipment availability, weather conditions, and time windows), coordinates data acquisition, preprocessing, and allocation of model computational resources, and can dynamically adjust schemes to respond to unexpected disruptions, ensuring the efficient advancement of scientific research tasks.

5) Embodied Intelligent Telescope: Physical Execution of Observation Tasks. As the physical execution layer of the system, this agent directly controls telescope hardware to perform operations such as pointing and tracking, filter switching, and exposure parameter adjustment. By integrating sensor data (such as guiding error and atmospheric seeing), it can monitor equipment status in real time and implement closed-loop corrections to ensure data acquisition quality. Simultaneously, it possesses the capability to conduct networked collaborative observations across multiple telescopes, laying the technical foundation for future multi-agent (multi-telescope) array collaborative observations.

6) Meta-Review Agent: System-Level Evaluation and Review. By analyzing the entire chain from hypothesis generation through simulation to observation and data, this agent evaluates whether the current hypothesis, simulation, and observations meet scientific requirements, and composes an analytical evaluation report. If empirical testing validates the current hypothesis and simultaneously satisfies the current intent requirements, the process enters the research summary phase; otherwise, it enters the reflective evolution phase.

7) Reflective Evolution Agent: Reflection on Factors Falling Short of Expectations and Iterative Optimization. Simulating the peer review mechanism, this agent conducts critical examination of scientific hypotheses, simulation screening, task embodiment, etc. Its evaluation dimensions include:

consistency with established physical laws, explanatory power for anomalous phenomena, verifiability, and degree of innovation. By comparing failure cases and successful paradigms from historical research, this agent identifies potential defects in hypotheses and generates improvement recommendations. Based on these recommendations, algorithms, models, or parameters are improved and upgraded through incremental learning incorporating new observational data or physical models—for example, methodology upgrades for flare eruption prediction, sunspot morphology classification, and magnetic field evolution prediction in the Jinwu large model.

8) Research Summary Agent: Composition of Research Reports and Whole-Process Experience Summarization. Reviewing the entire chain from intent parsing to meta-review, this agent composes a research report covering the research background and significance, attempted scientific hypothesis methods, simulation strategies, design and implementation of observation tasks, and the process of adjustment and upgrading. Simultaneously, it summarizes the successes and failures encountered during the process, providing reference for subsequent research.

Through the above model and agent design, SIDEST forms a workflow of Intent Parsing → Scientific Hypothesis → Simulation Screening → Task Embodiment → Embodied Observation → Meta-Review → Reflective Evolution / Research Summary. Within this workflow, these agents do not operate linearly but rather exist in a dynamic, mutually collaborative and competitive cycle. The agents challenge and refine each other's perspectives through a process akin to "scientific discussion," ultimately selecting the optimal hypotheses through ranking and evolutionary processes for simulation, subsequently driving telescopes to conduct real observational verification and continuously optimizing and iterating to achieve evolutionary autonomous research.

### 3.2.3 Self-Evolving Closed-Loop System

The SIDEST system adopts a three-level closed-loop architecture to achieve full-process autonomy from scientific intent to research evolution. The system comprises one system-wide closed loop and two internally nested closed loops, which together support the scientific research closed-loop evolution of SIDEST. The system-wide closed loop covers the complete scientific research chain from intent parsing to research summarization, encompassing the full workflow of Intent Parsing → Scientific Hypothesis → Simulation Screening → Task Embodiment → Embodied Observation → Meta-Review → Reflective Evolution / Research Summary. This loop emphasizes a "human-in-the-loop" interaction mechanism, ensuring that scientists can supervise and intervene in critical stages, thereby endowing SIDEST with the capability to distill new scientific questions from existing research and update research paradigms. Internal nested closed loop one targets the scientific hypothesis demonstration loop, conducting multiple rounds of feasibility demonstration of hypotheses through a simulation environment. This component replaces traditional manual derivation with computational simulation, rapidly screening for scientifically valuable and verifiable research hypotheses, thereby enhancing the rationality and realizability of scientific objectives and laying the foundation for subsequent observation tasks. Internal nested closed loop two targets the high-quality observation loop. During the task embodiment and embodied observation phases, the system dynamically adjusts observation strategies and equipment parameters by comparing real-time observational data with task objectives, enhancing the telescope's adaptive capability in variable observation scenarios and providing an extensible and reusable architectural paradigm for next-generation intelligent scientific observation facilities.

The core of the SIDEST system lies in its dynamic debate mechanism, multi-source knowledge fusion, and continuous evolution capability, enabling it to leverage the role of scientists in goal setting and outcome value judgment during the research process—ensuring scientific rigor while enhancing exploration efficiency, thereby forming a new AI-collaborative research paradigm of "scientific intent guidance + efficient machine execution."

### 3.3 System Implementation and Deployment

The deployment and implementation of the SIDEST system is a comprehensive endeavor integrating automatic control, robotics, AI technologies, and scientific research methodology. Its methodological core lies in a multi-agent architecture that simulates the observation planning and analytical reasoning processes of scientific research, achieving self-iterative optimization through techniques such as online reinforcement learning. This system decomposes complex scientific research tasks, which are then completed through the collaboration of multiple specialized agents. The basic process of system implementation and deployment is as follows:

1) Construction of Embodied Intelligent Telescopes: The embodied intelligent telescope is the foundation for realizing the current SIDEST. To validate the concept of an embodied intelligent telescope, we designed a master control system with perception, interaction capabilities, and automated observation execution for the Full-disk Vector Magnetograph (SFMM) of the Meridian Project Phase II, achieving intelligent operation in a "human-supervised, unmanned-operation" mode.

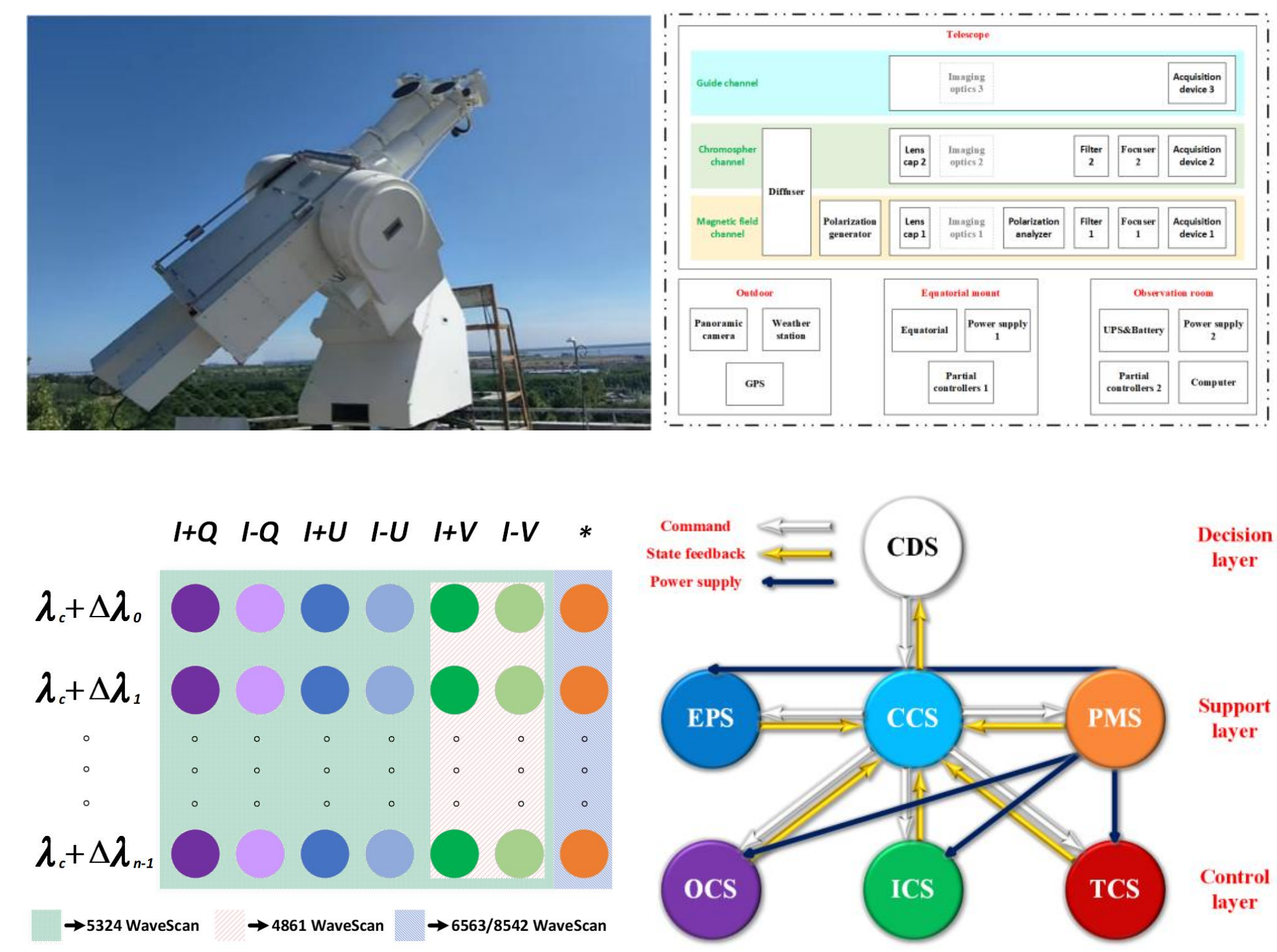

*Fig. 2 SFMM Master Control System*

Fig. 2 *a SFMM Telescope* *b The hardware composition of SFMM* *c Observation process of SFMM* *d SFMM's master control architecture*

In the SFMM master control system, an Environmental Perception System (EPS) was designed, such as the all-sky cloud cover monitoring and analysis system, which can achieve environmental perception based on current all-sky cloud cover and meteorological station data, providing environmental parameter support for system action decisions. Through systems such as the Telescope Control System (TCS), Instrument Control System (ICS), and Power Management System (PMS), the current operational status and parameters of the telescope are obtained, providing equipment parameter support for system driving decisions.

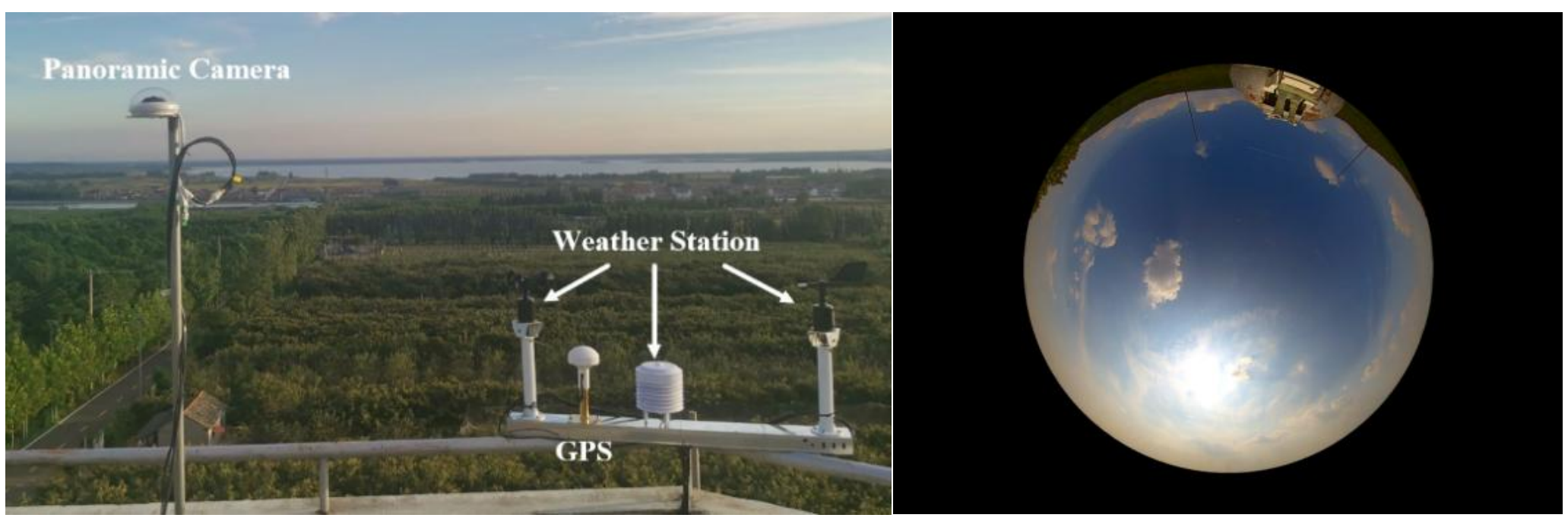

*Fig. 3 a SFMM's Weather Station and Panoramic Camera b SFMM's Total Cloud Amount Monitor*

A software system was developed based on a microservice architecture. Through the Communication and Control Subsystem (CCS), other subsystems are connected to the Central Decision Subsystem (CDS), which integrates observation tasks and evaluates telescope status (the right side of CDS displays the status of all telescope subsystems), achieving efficient system parameter monitoring and task scheduling.

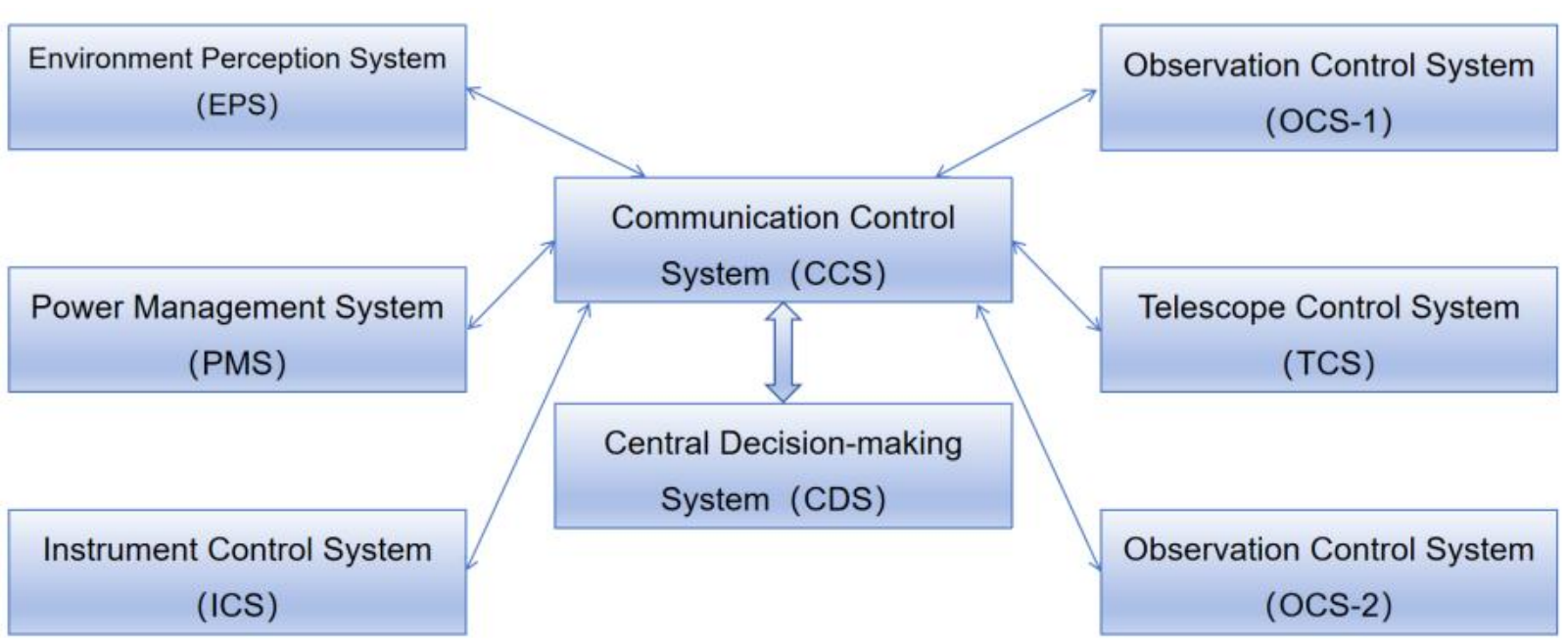

*Fig. 4 SFMM Subsystem Communication Diagram*

At present, this telescope has already acquired the capability to execute embodied observation tasks based on meteorological environmental conditions, telescope operational parameter perception, and observation target requirements.

2) System Design and Simulation Environment Construction: Construct a highly reliable, testable software and hardware foundational framework, define the system's autonomy level, select appropriate computing hardware and sensors, and rationally allocate computational resources. Design standardized communication interfaces to ensure that AI agents can seamlessly interact with the telescope control system (TCS/ICS), avoiding instruction conflicts or data packet loss. The simulation platform serves as the foundation for subsequent model training and validation, significantly reducing the risk of directly operating real equipment.

3) Model Training and Validation: On the one hand, utilize historical scientific data to construct foundational models oriented toward domain-specific scientific problems; on the other hand, employ the telescope system's historical observation logs and expert operational data to enable agents to master basic observation workflows through imitation learning. Further introduce reinforcement learning, establishing reward functions centered on data quality and observation efficiency to optimize multi-objective scheduling and anomaly response strategies.

4) Progressive Deployment Strategy: The deployment process must follow the progressive principle of risk-controllable advancement [24]. First, adopt a shadow mode in which the AI system receives telescope data in parallel and generates decision recommendations, but these are provided solely as reference for human operators without directly controlling hardware, thereby evaluating decision accuracy. Subsequently, enter a human-machine collaborative mode, authorizing the AI to autonomously execute low-risk tasks (such as flat-field and dark-field calibration), while requiring human confirmation for critical operations (such as target switching or anomaly interruption). Ultimately, after the system passes long-term stability testing, fully autonomous mode may be authorized, allowing it to independently complete the observation closed loop under preset conditions, with the human role transitioning to goal setting and result supervision. This process must establish clear rollback mechanisms to respond to unexpected failures.

5) Establish a Real-Time Closed-Loop and Continuously Iterating Evolutionary System: Once the system is formally operational, observation images and sensor data streams are continuously fed into the SIDEST system models, enabling dynamic adjustment of system models and agent parameters (e.g., adaptive exposure parameters based on seeing conditions). Based on an online learning framework, models are continuously optimized using new observational data and recent research literature, and new research reports are generated to summarize research outcomes and formulate scientific discoveries. System performance is periodically evaluated, and decision logic is iteratively upgraded in combination with feedback from human experts, ultimately forming an autonomous evolutionary loop of "observation–learning–optimization."

To achieve low-latency decision execution, some models should be deployed at the telescope side and must undergo compression, pruning, or distillation to accommodate the limited computational and storage resources of edge devices. Concurrently, to achieve real-time inference and ensure timely responses when the observation environment or targets evolve rapidly, efficient computational hardware (such as GPUs and FPGAs) and optimized software frameworks must be configured.

## 4 Experimental Design and Results

In the design of SIDEST, the main research content comprises three parts: scientific intent understanding and hypothesis demonstration, embodied intelligent observation, and evaluation and iteration. Among these, for the scientific intent understanding and hypothesis generation component, teams such as Google have already validated a series of feasible methods based on LLMs. For the embodied observation component, we have preliminarily implemented a viable foundational scheme on SFMM. At present, what still requires demonstration in the SIDEST system is the LLM-based capability for hypothesis simulation and result-based iterative optimization. To verify the technical pathways for these two components while avoiding the hypothesis space explosion and simulation infeasibility caused by complex scientific problems, we selected the relatively mature engineering domain to validate this design approach—specifically, designing an AI engineer in the domain of precision temperature control for exploratory experimentation. Precision temperature control is a critical foundational guarantee in fields such as scientific instrumentation (e.g., precision temperature control of solar telescope birefringent filters), advanced manufacturing (e.g., semiconductor lithography), aerospace, and bioengineering. Such systems typically exhibit highly nonlinear, large-inertia, pure time-delay, multivariable, strongly coupled, and easily perturbed complex dynamic characteristics. Traditional precision temperature control methods, such as PID (Proportional-Integral-Derivative) control and its variants (e.g., fuzzy adaptive PID, incremental PID), while performing well under specific operating conditions, often face bottlenecks in implementation: they require targeted research and debugging for different operating conditions, involve time-consuming learning processes, exhibit strong model dependence, and have limited adaptability and generalization capability.

To realize an LLM-based precision temperature control system with autonomous analysis and intelligent control capabilities, we proposed three progressively advanced, risk-controllable implementation approaches: (1) LLM as a tuner or optimization advisor: in offline or non-real-time scenarios, leveraging the LLM's powerful knowledge integration and reasoning capabilities to provide parameter optimization recommendations for traditional controllers (e.g., PID). (2) LLM as a high-level supervisory and decision-making layer: constructing a hierarchical control architecture in which the LLM is responsible for high-level strategy formulation, anomaly diagnosis, objective setting, and control strategy code implementation, while the underlying layer continues to execute real-time control via encapsulated MPC controllers. (3) End-to-end LLM controller: a more forward-looking approach in which the LLM directly serves as the controller (replacing conventional host computers such as microcontrollers or industrial PCs), receiving sensor data as input and outputting actuator commands to achieve adaptive control of complex, nonlinear systems.

The first approach cannot transcend the inherent limitations of the PID algorithm (e.g., certain operating conditions better suited to fuzzy control). The third approach remains controversial regarding whether the discrete, probabilistic output characteristics of LLMs can achieve long-term stable precision control. Therefore, in our experiments, we adopted the second approach—implementing an AI engineer system based on a subset of agents designed according to the scientific intent understanding and hypothesis demonstration concept in SIDEST, integrating the LangGraph workflow engine, RAG retrieval engine, MCP protocol manager, independent temperature monitoring, and other technologies. Its architecture is illustrated in Fig. 5.

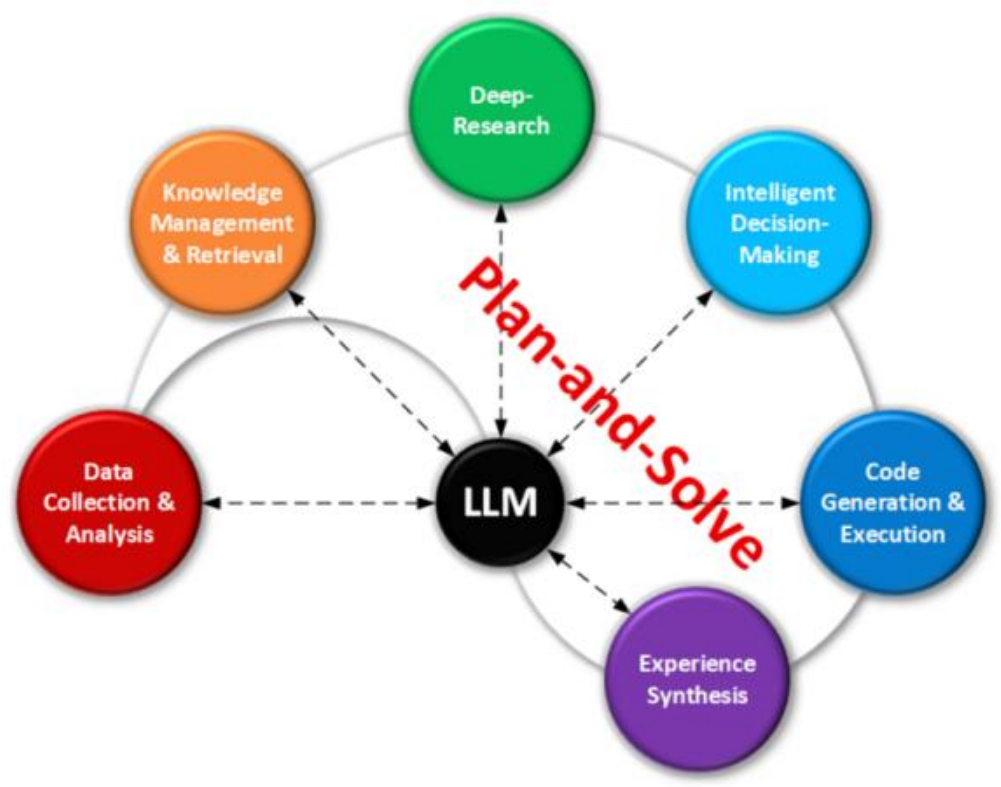

*Fig. 5 AI Engineer System Architecture Diagram*

### 4.1 Experimental Design

Based on the existing precision temperature control system of the birefringent filter of the full-disk solar magnetic field telescope at the Huairou Solar Observing Station, we performed adaptive modifications to meet the specific experimental requirements. The core modification consisted of disabling the closed-loop precision temperature control algorithm (e.g., integral-separation PI control) in the lower-level controller, retaining only high-precision temperature acquisition (using 24-bit AD), heating and temperature control drive, and necessary hardware protection mechanisms. Simultaneously, to ensure system safety, limit protection was implemented at the software level for the control quantities output by the LLM. The system architecture is shown in Fig. 6. The upper-level controller integrates an LLM-based hierarchical control architecture. Through the design of agent workflows, it achieves temperature data acquisition and characteristic analysis, literature review, in-depth research, and the composition of temperature control code on the upper-level computer, realizing closed-loop control through the driving interface and ultimately achieving the precision temperature control function of the filter.

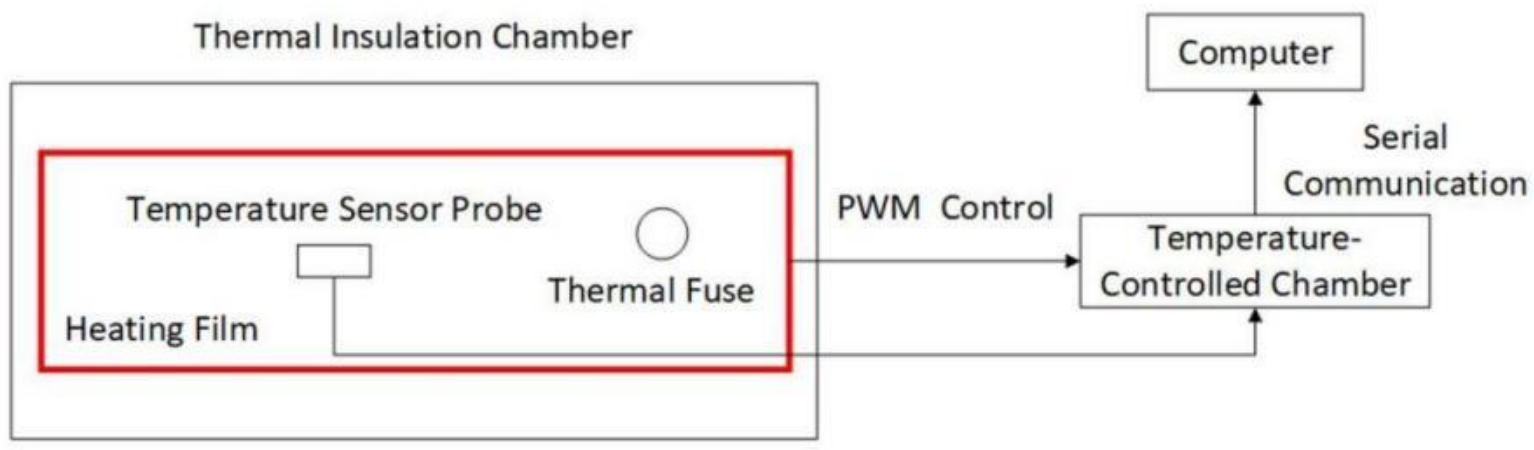

*Fig. 6 Precision Temperature Control System Diagram*

This experimental system is implemented through the combination of LLMs and agent workflows. The specific process is as follows:

Step 1: The System Cognition Agent acquires and controls data through hardware interfaces and analyzes data characteristics and variation trends.

Step 2: The Fundamental Knowledge Management Agent analyzes the local knowledge base to form a foundational control understanding and strategy for the system.

Step 3: The In-Depth Research Agent conducts deep investigation through the local knowledge base combined with web research to plan the latest control strategies for the system.

Step 4: The Control Strategy Agent, based on strategy templates, MCP extensions, and new strategies from in-depth research, formulates the current optimal control strategy.

Step 5: The Code Generation and Simulation Verification Agent, based on the optimal control strategy, conducts in-depth investigation and references open-source library materials from similar projects, generates code oriented to the current research topic in a sandbox environment, and executes code testing after safety verification.

Step 6: The Research Summary Agent, based on the current research, conducts experience summarization and reward scoring, and decides on further research strategies based on whether the control stability results have reached the expected level. If the target has been achieved, the current strategy is maintained and the strategy library is updated. If the target has not yet been reached, Steps 3 through 6 are repeated, realizing the self-evolution of system capability.

### 4.2 Experimental Results

Based on the AI temperature control engineer design described in Section 4.1, we conducted constant-temperature control tests on the solar telescope filter, with a target temperature of 32.7°C and an ambient laboratory temperature of approximately 25°C. The final temperature control results showed that the peak-to-peak stability precision of the control system was within ±0.0015°C (as shown in Fig. 7), meeting the requirements of the filter precision temperature control system. This result demonstrates that the AI temperature control engineer based on the SIDEST design approach can autonomously research—reviewing the literature, proposing constant-temperature control algorithms and software, and achieving the research objective of a precision constant-temperature control system through iterative refinement.

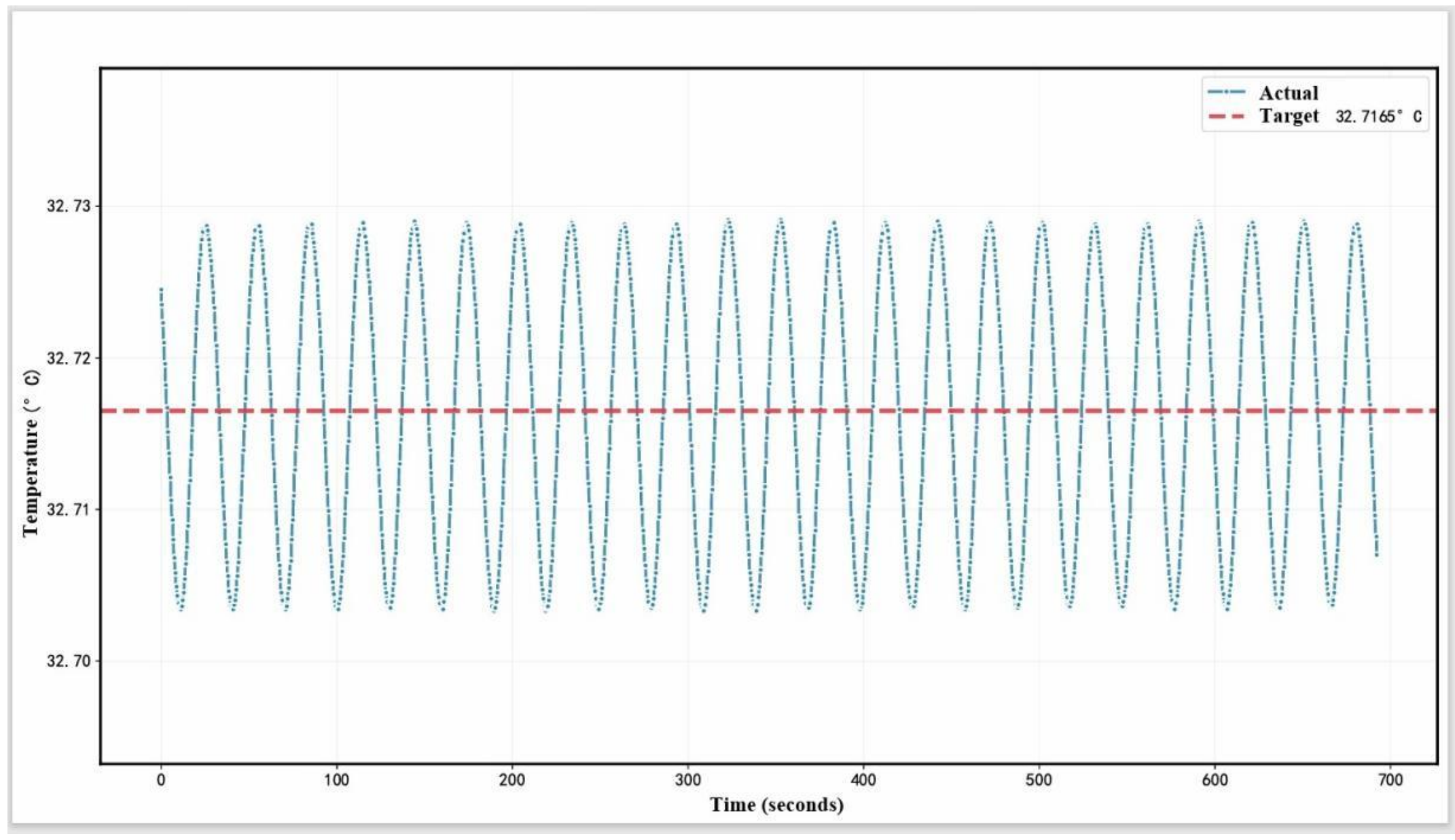

*Fig. 7 Results of Precision Temperature Control*

The system required only three weeks from initial design to achieving precision temperature control, whereas training a human engineer capable of developing precision temperature control system software typically requires several years. This fully demonstrates the high efficiency of the SIDEST framework in realizing an AI engineer and reveals the potential of this framework in scientific research. In the system design, we adopted a modular architecture, encapsulating different sensors and actuator hardware as unified MCP service modules, which allows rapid extension to other types of control systems—for example, motion control platforms—within the same framework. Therefore, this AI engineer system possesses good extensibility and generalization potential, and its prototype modules provide a reliable technical foundation for the implementation of the SIDEST system.

## 5 Summary and Outlook

The Scientific-Intention Driven Embodied Intelligent Solar Telescope (SIDEST) system proposes a new-generation telescope system architecture oriented toward autonomous scientific research. This paper has systematically elaborated its design philosophy and compositional structure, and validated the feasibility of the technical pathway through the construction of a prototype system for precision temperature control of solar telescope birefringent filters. Compared with existing autonomous research platforms, SIDEST's innovations are principally reflected in the following two aspects:

1) Scientific-Intention-Driven Software-Hardware Co-Design Architecture: The system takes the three solar telescopes at the Huairou Observing Station as its hardware foundation, constructing a telescope network capable of responding to a broad range of solar physics scientific intentions. At the software level, a three-level nested closed-loop agent architecture is designed to realize full-process autonomous research capabilities spanning from scientific intent parsing and demonstration, through observation task realization, to system self-evolution. The SIDEST architectural design supports intelligent networking and collaboration of multiple instruments, exploring technical pathways toward constructing an AI-driven, networked super-observatory (AI-empowered high-end scientific instruments [33], super-laboratory).

2) A New Paradigm of Deep Human-AI Collaborative Scientific Research: This system establishes a novel research modality of deep complementary collaboration between human scientists and AI scientists. Under this paradigm, human scientists are responsible for proposing key scientific questions and judging the value of research outcomes, while the AI scientist system undertakes highly repetitive, high-intensity tasks such as observation execution, data analysis, and model optimization—the two forming an efficient, synergistic closed loop. In the SIDEST system design, from intent input to result evaluation and throughout the intermediate processes, "human-in-the-loop" interaction interfaces are ensured, guaranteeing that human scientists can play the critical roles of goal setting and value judgment during the process. This paradigm preserves the dominant position of humans in creative thinking and value judgment in scientific research while leveraging the advantages of AI systems in data processing and pattern recognition, opening possibilities for more efficient scientific research and discovery.

Although the transition of this concept from theoretical conception to engineering practice still faces multiple challenges—for instance, in terms of system integration and performance optimization, the embodied intelligent system places extremely stringent demands on the stability, control precision, and response speed of telescope hardware. The complex astronomical observation environment requires the establishment of highly reliable software-hardware co-control mechanisms to ensure observation data quality and system robustness. At the model and knowledge level, the construction of domain-specific LLMs depends on large-scale, high-quality annotated data; however, annotation resources in the field of solar physics remain relatively scarce at present. Furthermore, how to effectively embed domain knowledge into models and ensure that their reasoning processes exhibit logical consistency, physical plausibility, and interpretability, while reducing hallucination phenomena, remains a key problem requiring in-depth exploration. Additionally, accurately delineating intellectual property attribution in automated scientific research is an important research ethics issue requiring attention. We believe that with the rapid development of AI technologies, the above issues will receive broader attention and sustained efforts, and the related technical challenges will be gradually identified and systematically resolved.

AI is profoundly transforming the paradigms of scientific research. We propose the SIDEST concept with the intention of redefining astronomical telescopes through AI from the perspective of technological development trends and exploring new paradigms for astronomical research. In the future, telescopes (and scientific instruments) will no longer be merely passive research tools that execute instructions, but rather research partners capable of understanding scientific objectives, proactively conducting exploration, and possessing continuous learning abilities. With the assistance of AI, the paradigm of scientific discovery will also gradually evolve from reliance on serendipitous inspiration and intuition toward a systematic, plannable "engineering" model. Its core lies in systematically identifying knowledge gaps, mining latent pattern combinations, and proactively targeting high-value breakthrough directions through large-scale historical data analysis and online experimental exploration. This transformation aims to construct a more efficient, reproducible, and continuously boundary-expanding new paradigm of scientific research [25]. Under this AI-driven new paradigm, the barriers to participation in scientific research will be substantially lowered. Scientific research will no longer be confined to the community of professional scientists who have undergone decades of specialized training; any individual with scientific curiosity and a spirit of exploration may potentially participate in, or even lead, research processes through natural language interaction [26]. This transformation will fundamentally broaden the scope of scientific research participants and drive a profound change in the research paradigm toward greater openness and collaboration.

## References

[1] Pesnell, W. D., Thompson, B. J., & Chamberlin, P. 2012, The Solar Dynamics Observatory (SDO) (Springer)

[2] Gan, W.-Q., Zhu, C., Deng, Y.-Y., et al. 2019, Advanced Space-based Solar Observatory (ASO-S): an overview, Research in Astronomy and Astrophysics, 19(11), 156

[3] Huang, K., Hu, T., Cai, J., Pan, X., Hou, Y., Xu, L., Wang, H., Zhang, Y., & Cui, X. 2023, Intelligence of Astronomical Optical Telescopes: Present Status and Future Perspectives, Journal of Astronomical Instrumentation, 12(4), 1–45

[4] Tong, L. Y., Sun, Y. Z., Yang, X., et al. 2024, Design and application of an autonomous Master Control System for a multi-layer magnetic and helioseismic telescope, Astronomical Techniques and Instruments, 1(3), 1–10

[5] Bloom, N., Jones, C. I., Van Reenen, J., et al. 2020, Are ideas getting harder to find?, American Economic Review, 110(4), 1104–1144

[6] Frank, M. C. 2023, Baby steps in evaluating the capacities of large language models, Nature Reviews Psychology, 2(8), 451–452

[7] Brown, T., Mann, B., Ryder, N., et al. 2020, Language models are few-shot learners, Advances in Neural Information Processing Systems, 33, 1877–1901

[8] Jiang, L. Y., Liu, X. C., Nejatian, N. P., et al. 2023, Health system-scale language models are all-purpose prediction engines, Nature, 619(7969), 357–362

[9] Singhal, K., Azizi, S., Tu, T., et al. 2023, Large language models encode clinical knowledge, Nature, 620(7972), 172–180

[10] Mirza, A., Alampara, N., Kunchapu, S., et al. 2024, Are large language models superhuman chemists?, arXiv:2404.01475

[11] Kramer, S., Cerrato, M., et al. 2024, Automated scientific discovery: from equation discovery to autonomous discovery systems, arXiv:2403.12345

[12] Silva, R. G. L. 2023, The advancement of artificial intelligence in biomedical research and health innovation: challenges and opportunities in emerging economies, Globalization and Health, 19(1), 45

[13] Gottweis, J., Weng, W. H., Daryin, A., et al. 2025, Towards an AI co-scientist, arXiv:2502.18864

[14] Wang, L., Ma, C., Feng, X., et al. 2024, A survey on large language model based autonomous agents, Frontiers of Computer Science, 18(6), 186345

[15] Kulkarni, A., Alotaibi, F., et al. 2023, Scientific hypothesis generation and validation: methods, datasets, and future directions, arXiv:2310.12345

[16] Ai, G.-X., & Hu, Y.-F. 1986, The Proposal for the Solar Magnetic Field Telescope and Its Working Theorem, Acta Astronomica Sinica, (2), 91–98

[17] Zhang, H. Q., Wang, D. G., Deng, Y. Y., et al. 2007, Solar Magnetism and the Activity Telescope at HSOS, Chinese Journal of Astronomy and Astrophysics, 7(2), 281–288

[18] Djorgovski, S. G. The Roles of Small Telescopes in a Virtual Observatory Environment, Astrophysics and Space Science Library

[19] Li, Y. Y., Bai, Y., Wang, C., et al. 2024, Deep Learning and LLM-based Methods Applied to Stellar Lightcurve Classification, Intelligent Computing, doi:10.34133/icomputing.0110

[20] Bu, K., Liu, Y. C., Ju, X. L. 2024, Efficient utilization of pre-trained models: A review of sentiment analysis via prompt learning, Knowledge-Based Systems, 283, 111148

[21] Elkhatib, Y., Coulson, G., Tyson, G. 2017, Charting an intent driven network, In: Proc. of the 2017 13th Int'l Conf. on Network and Service Management (CNSM), IEEE Computer Society

[22] King, R. D., Whelan, K. E., Jones, F. M., et al. 2004, Functional genomic hypothesis generation and experimentation by a robot scientist, Nature, 427(7059), 247–252

[23] Lin, J. B., Shen, Y. B., Zhu, X. M., et al. 2013, Design of the Automatic Observing System for Full Disk Magnetogram in HSOS, Astronomical Research and Technique, 10(4), 392–396

[24] Wang, C. S., Hu, X. J., Zhang, Y., et al. 2024, StarWhisper Telescope: Agent-Based Observation Assistant System to Approach AI Astrophysicist, arXiv:2412.06412

[25] Liu, Z. Y. 2025, The Fifth Paradigm, Citic Press Corp., Beijing

[26] Mao, Y. N., Xiang, E., Li, Y. Y., Wang, C. S., Liu, J. F. 2025, Artificial Intelligence-Driven Construction and Practical Exploration of an Astronomical Science Education System, Science Education for Primary and Secondary Schools, 2(2), 41–46

[27] Mitchener, L., Yiu, A., Chang, B., et al. 2025, Kosmos: An AI scientist for autonomous discovery, arXiv:2511.02824

[28] Swanson, K., Wu, W., Bulaong, N. L., Pak, J. E., Zou, J. 2025, The Virtual Lab of AI agents designs new SARS-CoV-2 nanobodies, Nature, 646(8085), 716–723

[29] Tang, J., Xia, L., Li, Z., & Huang, C. 2025, AI-Researcher: Autonomous Scientific Innovation, arXiv:2505.18705

[30]Institute of Artificial Intelligence and Robotics, Xi'an Jiaotong University. 2025, http://www.aiar.xjtu.edu.cn/info/1004/3785.htm

[31] Shao, M., Liu, S., Xu, H., Jia, P., Wang, H., Tong, L., Bai, Y., Yang, C., Li, Y., Li, N., & Lin, J. 2025, Advances and Challenges in Solar Flare Prediction: A Review, arXiv:2511.20465

[32] Bai, C. L. 2025, AI Empowers High-end Scientific Instruments to Drive the Leap-forward Development of Intelligent Scientific Research, World Sci-Tech R & D, doi:10.16507/j.issn.1006-6055.2025.11.001

# 科学意图驱动的具身智能太阳望远镜：概念设计

林佳本 [1]，佟立越 [1]，王慧 [1,2]，邵明福 [1,2]，杨晨 [1,2]

[1] 太阳活动与空间天气全国重点实验室，中国科学院国家天文台，北京，100101

[2] 中国科学院大学，北京，100049

**摘 要** 人工智能（AI）正在深刻变革科学研究的范式，大语言模型、具身智能等前沿技术正在助力科学仪器不断突破其原有的能力边界。在此背景下，本文提出了科学意图驱动的具身智能太阳望远镜阵（Scientific-Intention Driven Embodied Intelligent Solar Telescope，SIDEST）这一新的概念系统，该系统设计了三个核心层次实现三种类型的智能化科学研究闭环：首先，科学意图研究论证层通过自然语言交互解析科学家（如太阳物理学家）的研究目标意图，通过深度研究实现与科学意图相匹配的可执行观测方案的生成和优化闭环；然后，观测具现层调度具身智能太阳望远镜，实现科学观测计划落地执行闭环；最后，评估演化层协调数据处理与科学分析智能体对观测数据进行分析、生成研究报告，并基于结果迭代优化观测策略与模型方法，实现全系统的自我演进闭环。研究过程中，我们基于太阳望远镜滤光器精密温控装置构建了一个试验 SIDEST 原理的最小原型系统，实现了面向研究意图的自动化科研的所有关键步骤，验证了三类智能化研究闭环技术路径的可行性。SIDEST 用前沿的 AI 方法重新定义望远镜，通过融合 AI 的推理、规划能力和科学家的创造性思维，构建"人机协同"的天文研究新方法，为下一代天文望远镜、智能仪器的研制和天文研究新范式探索可能的发展方向和可行的技术路线。

# Scientific-Intention Driven Embodied Intelligent Solar Telescope: Conceptual Design

Lin Jiaben[1*], Tong Liyue[1], Wang Hui[1,2], Shao Mingfu[1,2], Yang Chen[1,2]

[1] *State Key Laboratory of Solar Activity and Space Weather, National Astronomical Observatories, Chinese Academy of Sciences, Beijing, 100101, China*；

[2] *China University of Chinese Academy of Sciences, Beijing 100049, China;*

**Abstract** Artificial Intelligence (AI) is profoundly transforming the paradigms of scientific research. Cutting-edge technologies such as Large Language Models and embodied intelligence are continuously pushing the boundaries of scientific instrumentation. Against this backdrop, this paper proposes a novel conceptual system: Scientific-Intention Driven Embodied Intelligent Solar Telescope (SIDEST). The system is designed with three core layers to achieve three types of intelligent scientific research loops. First, the Scientific Intent Research and Demonstration Layer parses the research objectives and intents of scientists (e.g., solar physicists) through natural language interaction. It achieves a closed loop involving the generation and optimization of executable observation plans that align with the scientific intent via in-depth research. Subsequently, the Observation Realization Layer schedules embodied intelligent solar telescopes to implement a closed loop for the execution of scientific observation plans. Finally, the Evaluation and Evolution Layer coordinates intelligent agents for data processing and scientific analysis to analyze observation data, generate research reports, and iteratively optimize observation strategies and model methods based on the results, thereby realizing a self-evolving closed loop for the entire system. During the research process, we developed a minimal prototype system based on a precision temperature control device for solar telescope birefringent filters to validate the core principles of SIDEST. This prototype successfully implemented all key steps of intention driven automated research, demonstrating the feasibility of the technical pathways for the three types of intelligent research loops. SIDEST aims to redefine solar telescopes through cutting-edge AI methods. By integrating AI's reasoning and planning capabilities with scientists' creative thinking, it seeks to establish a new "human-machine collaborative" approach for solarphysics research. This initiative explores potential development directions and viable technical pathways for next-generation astronomical telescopes, intelligent instrumentation, and new paradigms in astronomical research.

## 1 引　言

太阳是太阳系的中心天体，主宰着太阳系的演化进程，为地球生命提供不可或缺的能量来源。它不仅是天体物理学研究的关键样本，也是空间天气监测和预报中需要持续关注的风暴源头。随着现代高科技系统的日益发展和向太空的拓展，太阳活动——尤其是太阳风暴——对人类技术社会的潜在破坏性影响愈发凸显。为应对这些可能的影响和挑战，世界各国已部署了大量天基和地基太阳监测设备，如太阳动力学观测台（SDO）[1]和“夸父一号”[2]（ASO-S，先进天基太阳天文台），这些观测设备推动太阳观测进入每天 TB 级的海量数据时代，为科研人员提供了前所未有的研究机遇。虽然观测技术、数据处理分析技术都已实现跨越式发展，但是当前的天文学究方法（包括太阳物理研）在本质上仍延续着伽利略时代以来的天文学研究范式：由天文学家提出假设、设计观测方案[3-4]，并通过人工分析数据来推动科学发现。这种研究方式高度依赖个人经验与直觉，其效率和科学产出能力已难以适应当今海量数据和复杂科学问题的要求。同时，相关研究表明，在这类科研范式下，多数学科领域的进步速率大约每 13 年就会降低一半[5]。由此可见，无论是从太阳物理、空间天气学科研究的需求，还是从科研方法本身的发展趋势来看，都迫切需要新的范式来进一步提高研究效率和能力。

近年来，以大型语言模型（Large Language Models, LLMs）为代表的人工智能技术在知识理解、任务规划与逻辑推理等方面展现出显著能力[6-10]，不断地与各个学科领域结合，成为科学研究中的重要工具。以通用语言大模型为基础，化学、生物医药等很多学科的研究者们都构建了自己专业领域的大模型、智能体系统，这些系统具备了自主完成“提出假设-实验验证-结果分析”全流程的能力，从而构成真正意义上的自动化科研系统[11-13][27-29]。但是，这些系统主要是针对确定类型的科学问题开展研

究工作，是否能够让系统理解研究者以自然语言形式输入的研究意图，并自动化开展宽领域、多方向的研究？

实际上，大语言模型的发展为为意图识别任务提供了一种高效的技术路线——通过微调基础模型、设计专业提示词以及检索增强等方法可以实现高准确率的意图识别。与此同时，以大模型为驱动核心的具身智能（Embodied AI）也在快速发展，当这一概念应用于机器人、实验科学中，便催生了能够实现自动化开展试验的智能仪器系统。此类系统可以不再局限于被动执行预设指令，而能够通过与环境、试验条件的交互理解实现高效的试验、研究[14]。例如，在生物化学与新材料研究领域，通过整合机器人实验平台、多模态感知模块、自适应控制算法以及大模型的认知推理能力，构建出可连续运行的自动化研究系统，显著提升了新材料筛选与化学反应路径优化的效率[15]；在天文领域，StarWhisper Telescope（SWT）系统基于大语言模型（LLM）构建智能代理系统，实现了天文观测的自动化管理，应用于近邻星系超新星巡天（NGSS）项目，整合了观测计划生成、控制、数据处理和实时反馈等功能，通过自然语言交互降低操作门槛，有效减轻天文学家的工作负担[24]。这类系统代表了科学研究范式的重要演进方向——即 AI 驱动的自主科研，也体现了多智能体协作在实验环境中的实际应用潜力。

基于上述技术发展背景分析，我们提出了“科学意图驱动的具身智能太阳望远镜”（Scientific-Intention Driven Embodied Intelligent Solar Telescope，SIDEST），这一面向下一代太阳物理研究的智能观测系统，其核心思想是赋予望远镜系统理解科学意图的能力，使其能够将研究者以自然语言表述的科学问题（如“研究太阳耀斑触发机制”、“分析活动区磁场演化与暗条演化的关联”等）自动转化为可执行的观测任务，并实现从任务规划、观测驱动、数据采集到结果分析的全流程研究闭环。SIDEST 系统将依托国家天文台怀柔太阳观测基地现有望远镜设施进行构建（包括一台 35 厘米太阳磁场望远镜[16]、一台全日面矢量磁场望远镜以及一台全日面色球（Hα）望远镜[17]，每台望远镜都进行具身智能化改造升级），通过深度融合大语言模型（LLMs）、多模态感知、自适应控制、深度研究等人工智能技术，形成一个具备自主观测、动态决策与持续演化能力的智能观测研究系统。该系统不仅能够识别、响应天文学家提出的科学目标，还能在观测过程中根据实时观测环境、望远镜运行状态和观测目标的信息反馈自动优化观测策略，开展对太阳活动现象的智能追踪和深度研究，实现天文观测研究范式从“人工驱动”向“人机协同”的智能化转变。

论文内容安排，第二节深入探讨科学意图驱动的具身智能太阳望远镜（SIDEST）概念的基本内容；第三节介绍 SIDEST 的基设计本框；第四节介绍其原型系统试验和结果；最后，探讨和展望这一概念未来的发展方向。

## 2 科学意图驱动的具身智能太阳望远镜概念

望远镜是人类探索宇宙的历程中不可替代的关键设备，它拓展了人类的感知与认知能力。长期以来，望远镜都是作为一种被动的执行工具，严格遵循天文学家与工程师预设的流程执行动作并产出科学数据。近年来，随着自动控制、机器人等技术的快速发展，尤其是以大模型与智能体为代表的人工智能技术的突破，天文望远镜同其他科学仪器一样，正经历着一场范式转变——向具身智能演进。在这一新范式下，望远镜（科学仪器）将不再是被动执行指令的装置，而是能够主动分析任务与环境、通过物理交互实现更优执行方案的智能机器[18]。其实，早在 1950 年图灵就提出了具身智能的概念，强调智能需通过“具身”与环境的动态交互实现自主学习和进化，其核心在于将感知、行动与认知深度融合。将这一理念应用于天文观测设备，便形成了具身望远镜概念。此类望远镜具备内外双重感知能力：对内可实时监测望远镜本体的指向误差、温度变化、仪器参数等全链路状态信息；对外则能精确感知气象条件、云量分布等可能影响观测的外部环境信息。在此基础上，通过深度融合人工智能技术，实现了从具身到智能的跃升，从而诞生了具身智能望远镜。它不再仅仅是自动执行预设方案的设备，而是能够根据实时环境条件、仪器状态和科学任务要求，自主进行决策、规划并执行观测任务的高级智能系统。

在具身智能望远镜的理论与实践基础上，本研究进一步提出了“科学意图驱动的具身智能太阳望远镜”这一概念系统。该系统基于前沿人工智能技术，旨在实现对太阳望远镜观测能力的系统性提升。其核心创新在于，突破了传统具身智能主要关注与物理环境交互的局限，通过引入对科学家研究意图的识别、理解与逻辑论证，实现了观测执行与科学研究意图之间更高层级的交互与协同演进，从而将观测设备从被动的数据采集工具转变为主动的科学发现伙伴。

SIDEST 的系统设计架构可抽象为三个核心层次，构成一个完整的感知、决策、行动、评估闭环：

科学意图研究论证层，该层对应从科学意图理解到可验证科学假设的生成与论证过程。作为 SIDEST 系统的出发点，依托大语言模型对自然语言观测请求的深度理解能力，构建能够理解科学家研究意图的智能系统，通过自然语言处理与知识图谱技术解析科学问题[19][20]，生成可执行的研究假设，并将其转化为观测目标与规划方案，实现观测规划的优化闭环。传统科研流程中，将科学意图转化为可执行方案通常需科学家与工程技术人员耗费大量时间，论证一系列可能支持该意图的假设空间解的可行性，并通过模拟仿真验证其有效性，之后方可制定具体观测计划。天文学作为高度交叉学科，常需多领域专家协作论证潜在假设，过程复杂耗时，SIDEST 则借助大语言模型强大的推理与规划能力，能够显著加速从研究意图到高价值假设筛选的实现。

观测任务具现层，该层对应观测任务中的工具调用、驱动和任务实现。它接收来自上一层的具体观测指令，动态调度并控制望远镜的硬件系统，在复杂多变的环境约束下，自主完成观测任务。该层依托大语言模型所具备的推理、规划与工具调用能力，结合望远镜具身智能系统控制框架（该框架将圆顶、指向系统及成像设备等硬件组件及相关软件工具封装为可编程技能库），形成从语义解析到任务编排的完整闭环。系统能够依据观测任务目标、实时天气条件及设备可用状态，动态调整观测优先

级，并生成符合规范的指令包下发至望远镜控制系统。在 SIDEST 的技术架构中，通过协同调度接口与可用望远镜网络对接，实现跨站点、多波段的联合观测智能协调。该协同机制对太阳爆发活动等现象的观测分析尤为重要，因为此类现象的研究通常依赖于多设备、多波段的高分辨率连续观测数据，才能够捕捉其完整的物理演化过程，而 SIDEST 能够根据科学目标与实时环境条件动态分配观测任务、调度设备，优化设备利用效率，从而实现更高时空分辨率的观测。

评估演化层：该层对应科学研究的反馈闭环与系统自主迭代。它对采集到的数据进行分析，评估其是否满足科学意图论证层提出的假设与需求，并将评估结果（如数据质量、任务完成度、新发现线索）反馈给系统。基于反馈，系统可以自动优化观测策略，调整科学假设，甚至提出新的探索方向，从而实现整个系统的持续学习和自主演化。SIDEST 的目标是构建具备自主科学发现能力的太阳望远镜系统，通过在线强化学习等方法对实测数据进行科学分析，并与科学意图进行比对，通过反复迭代优化实现高效的科学探索发现。例如，在耀斑监测任务中，系统将观测结果反馈至预测模型，通过对比分析生成科学发现，进而提升后续预测准确性，相关发现以标准化报告形式提交科研团队，同时更新系统知识库，形成从科学仪器到数据再到新的知识和发现的闭环。

# 3 系统架构与关键技术

SIDEST 设计了三个核心层次的框架结构，实现三类智能化科学研究闭环。其中，意图研究论证层作为系统与外界交互的接口，负责接收并解析高层次的研究目标或科学意图并逐步步分解为可论证的假设。观测具现层则承担将假设具现为具体观测任务的执行功能，确保科学数据的高质量采集与标准化输出。评估演化层作为系统的核心评估单元，负责统筹科学意图与观测数据，构建从预测判断到结果反馈的闭环机制，推动研究过程的持续迭代与系统能力的自主演进。系统架构如图 1 所示：

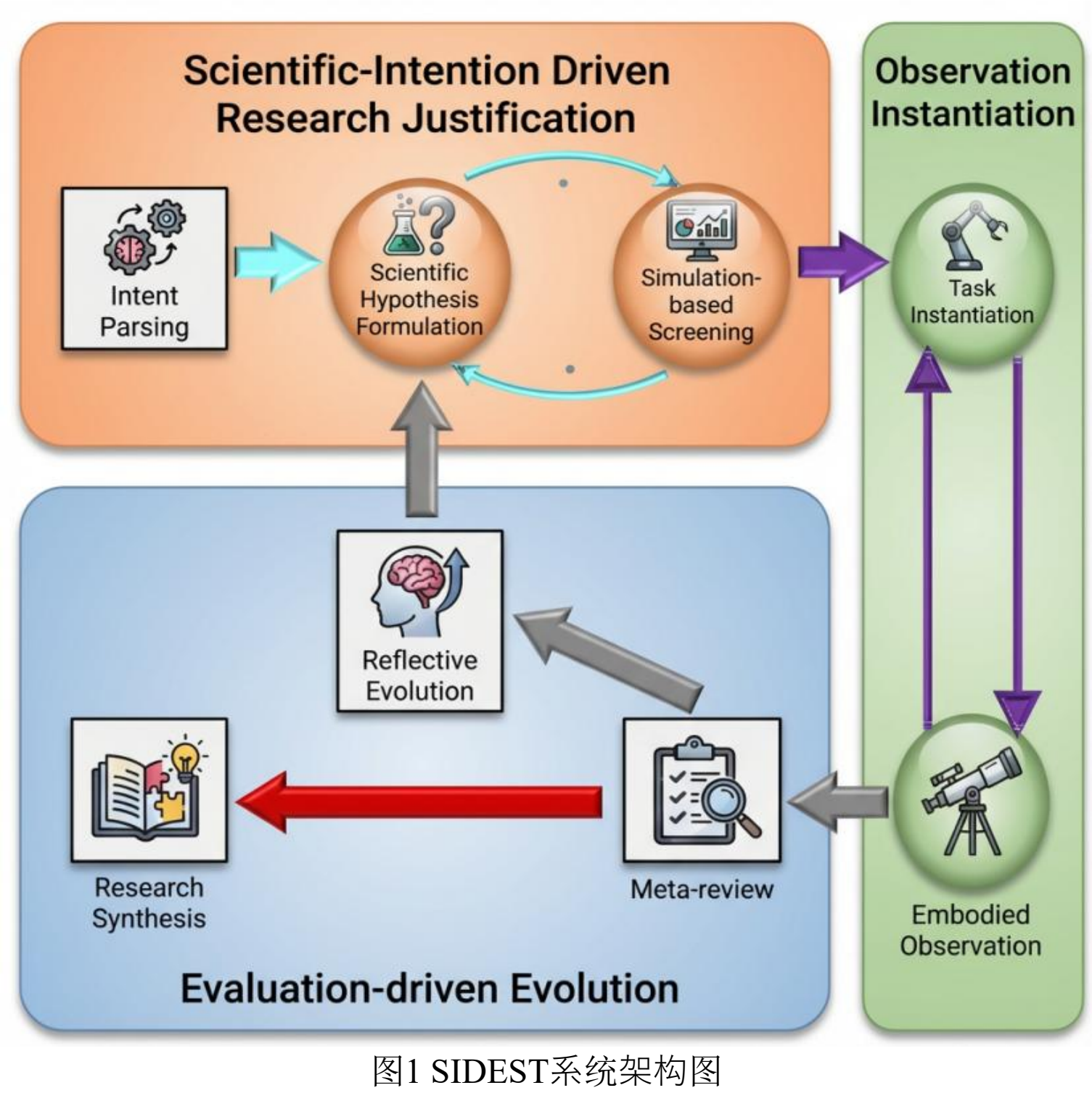

图1 SIDEST系统架构图
Fig. 1 SIDESTASystem Architecture Diagram

### 3.1 系统整体框架设计

本系统通过三层架构设计实现从科学意图解析优化、具身智能观测与科学研究的三重闭环。各个层次的核心功能与实现路径如下：

#### 3.1.1 基于大模型的科学意图识别与研究论证

意图驱动[21]概念最初源于网络通信领域，其目标是将用户以自然语言描述的业务需求自动转化为网络配置策略，从而实现网络的自动化运维。随后，该理念逐步拓展至人工智能领域，强调系统应具备对人类目标的深层理解能力，并能够自主规划行动路径，进而成为实现高度自治系统的核心范式。2025 年，郑南宁院士在 WAIC 大会上进一步提出“智能的跃迁：从模型驱动到意图驱动”的理论框架[31]，本文在上述研究的基础上进一步提出科学意图驱动的概念并将其延伸至太阳望远镜研制领域，形成了 SIDEST 概念系统。

在 SIDEST 系统设计中，首先构建以基础大模型为核心的科学意图接口层，支持科学家通过自然语言、图形界面或领域专用语言描述科学目标，例如：“基于活动区图像、磁场强度与中性线长度等参数，预测耀斑爆发时间与强度[32]”。其次，设计意图翻译与分解引擎层，该层由大语言模型、智能体和天文学知识图谱共同构成，通过多智能体驱动生成大量可论证假设，结合各项预设工具模拟仿真，论证可行的假设输出给下一层级的驱动接口。

#### 3.1.2　从观测方案设计到调度执行的具身智能望远镜

具身智能的概念在机器人领域有广泛的研究和应用，将这一技术融入传统天文观测设备，可构建具备“感知-理解-决策-行动”闭环能力的具身智能观测系统，该系统能够接收观测计划，并将其解析为望远镜指向、跟踪、对焦、观测终端切换与参数配置等底层控制指令；同时，借助分布于望远镜各部组件的传感器（如导行、位置、温度传感器等），实时感知仪器状态并形成反馈，确保执行过程的精确性与可靠性，从而在硬件层面实现物理闭环控制。

SIDEST 在具身智能望远镜的基础上，进一步提升了感知理解与规划能力。在获得可论证的假设后，系统首先分析实现假设所需的相关条件，包括天气条件、设备能力及当前可用设备与网络是否满足需求；继而提出可观测验证的假设，并将其具现为所需设备参数与可能的执行步骤；进而设计适配不同假设的实验方案，动态加载并调整观测任务列表，调度不同优先级与条件的观测计划至最优状态，最终实现观测任务的闭环执行。

#### 3.1.3　演进式自主化科学研究

自主化科学研究是指利用人工智能技术（特别是大语言模型与机器人系统）实现全科研流程自动化的研究范式，其核心在于通过 AI 智能体自主完成从假设生成与仿真论证、实验设计、观测调度执行、数据分析评估、实验迭代、论文撰写的完整循环，大幅减少人工干预。这一概念可追溯至“机器人科学家”与“自主科学发现”等早期探索[22]，其目标在于突破人类认知瓶颈，显著加速科学发现进程。

在 SIDEST 系统中，通过 AI 科学家/太阳物理学家实现演化式的自主化科学研究，例如，耀斑预测模型、智能体可以基于实时磁场数据预测耀斑爆发与否，调度具身智能望远镜开展有目的的观测；观测结果作为反馈信号，用于优化预测模型与方法——若爆发发生则强化现有模型，若未发生则触发模型重构甚至大模型重训练，从而持续提升预测准确性，该过程完整体现了“预测-行动-观察-学习”的演进式科学探索闭环。

SIDEST 通过融合科学意图理解、观测具现与演化式自主科研三层闭环能力，构建出能够协同开展太阳物理研究的智能太阳观测系统。该系统能够开展太阳磁场观测、活动爆发预测及色球爆发监测，形成观测-预测-监测-反馈的闭环逻辑，并通过在线强化学习实现面向科学意图的演进式自主化科学研究闭环。

### 3.2　关键技术、模块：仪器与数据、模型、系统

SIDEST 的实现需要仪器数据、模型算法与系统架构三个维度的关键技术支撑。仪器数据层面，需要构建长周期、高质量的科学数据集，并经过严格的标准化预处理，以保障模型构建的可靠性。同时，系统应具备持续采集新数据并生成 AI 就绪（AI-ready）处理结果的能力，以支持模型的持续演

进；模型层面，需要实现数据驱动与领域知识嵌入的深度融合，使模型或智能体既深刻理解领域知识，又适配科学数据的特性。模型训练与智能体配置需具备灵活性，以便有效集成与利用开源基础大模型的先进技术；系统层面，需构建多智能体协同的闭环框架，实现从科学假设生成、观测方案设计、流程自主执行到结果分析与优化迭代的全流程自动化；同时，系统各个环节应具备便捷的人机交互接口，确保科学家能够以“人在回路”（Human-in-the-loop）的方式参与并监督科研过程。

#### 3.2.1 关键仪器与数据基础

SIDEST 的研究中以怀柔基地现有望远镜、数据为基础，利用历史数据构建系统的初始算法、模型，在实际观测中自主演化提升系统算法、模型的性能指标。

1）矢量磁场望远镜与长周期太阳活动数据

怀柔观测基地的 35 厘米太阳磁场望远镜（1984 年建成）与太阳光学和磁活动监测望远镜（2006 年投入运行）长期以来一直保持着稳定可靠的观测，积累了覆盖多个太阳活动周的高精度光球与色球矢量磁场数据。目前，我们正在利用这些地面观测数据正与太阳动力学天文台（SDO）的日震与磁像仪（HMI）以及先进天基太阳天文台（ASO-S）的全日面矢量磁像仪（FMG）等空间仪器所获数据进行系统化的一致性定标。通过联合处理不同来源的磁场观测资料，将构建时间跨度达数十年的超长期太阳活动数据集。这些仪器数据集不仅可为太阳物理模型提供高质量的训练样本，其实时观测数据流还可作为在线强化学习的输入，驱动模型动态响应太阳活动的瞬时变化，为太阳活动预报与空间天气预警提供关键支持。

2）色球望远镜与爆发活动特征数据

怀柔太阳观测基地的全日面色球（Hα）望远镜自 2006 年投入运行以来，作为全日面磁场与活动监测望远镜（SMAT）的重要组成部分，持续提供高质量的太阳色球层观测数据。该望远镜与同平台安装的全日面矢量磁场望远镜协同工作，能够在 6562.8 埃波段获取全日面 Hα单色像，实现光球矢量磁场与色球活动现象同步观测[23]。其观测数据对太阳耀斑、暗条爆发和日珥活动等关键过程的识别与分类具有重要价值。目前，基于该望远镜积累的长期观测资料，SIDEST 研究团队正通过将其与 GOES 卫星的 X 射线数据进行交叉检验与标注，构建专门针对 Hα耀斑活动的标注数据集。这一数据集将显著提升对太阳爆发事件的识别可靠性，并为太阳活动监测系统提供反馈信号。在此基础上，该数据集可进一步集成至 SIDEST 系统中，支撑其实现观测、预测、监测、反馈优化的闭环控制流程。

3）太阳物理文献与教材的领域知识库

为实现对大模型“幻觉”现象的有效控制，提升其在专业领域内的知识一致性，SIDEST 研究团队对太阳物理领域的数百部经典教材与数千篇权威文献进行了系统性梳理与整合，构建了结构化的领域知识库。在此基础上，引入检索增强生成技术框架，使模型在生成回答前能够实时检索知识库中的相关知识作为依据，增强输出内容的科学准确性与可靠性。在“金乌”系列模型的实践中，该方法不仅有

效提升问答任务的专业可信度，而且能够缓解因训练数据更新滞后导致的领域知识时效性不足问题。进一步地，SIDEST 研究将通过持续扩展知识库规模并引入多模态标注（如因果链、时间线等结构化语义信息），提升模型的逻辑推理与复杂场景适应能力，以支撑高要求的科研推理任务。

4）具身智能望远镜

具身智能望远镜是一种集成了多模态感知、自主决策与闭环执行能力的新一代天文观测系统[30]。SIDEST 系统中，计划采用的太阳望远镜已经初步具备具身智能的能力。其中，外部环境感知由观测站配置的小型气象站、全天相机共同组成，能够根据外部的气象条件决定是否启动或停止观测；能够基于气象站数据和当前全景云量分析预判观测状况，提前做好观测规划；而望远镜内部运行状态，则由广泛分布于望远镜各个子系统的传感器实时反馈，如望远镜的指向位置、滤光器温度、滤光片位置、系统运行的电压电流状态等。一方面能够评估系统当前的观测能力，如望远镜方位、高度、各个仪器组件状态是否满足观测所需；另一方面也能评估各个观测模式下自身储能、安全状况，确保规划的观测任务能够按计划完成。

### 3.2.2 模型与智能体构建

SIDEST 系统通过一组由监督智能体协调的专业模型、智能体群组成，模拟科学家开展研究工作。这些智能体基于意图理解规划、观测执行与反馈迭代三大大环节构建闭环科研范式，其具体功能规划如下：

1）意图解析智能体：科学意图的感知与解析

该智能体作为系统输入接口，负责通过自然语言处理技术解析科学家以文本或语音提出的研究目标（如"研究太阳耀斑爆发的触发机制"）。它能够识别指令中的关键参数、隐含约束及优先级，并将其转化为结构化、可操作的研究任务描述。其核心能力包括多轮对话上下文维持、领域术语消歧及意图置信度评估，为后续任务分解提供精确的输入基础与背景调研。

2）科学假设智能体：假设创新与跨领域融合

通过整合网络搜索、文献数据库与领域知识库，该智能体能够生成初步研究假设与实验方向，并跨领域丰富假设内容。例如，针对太阳耀斑预测任务，它可以综合历史爆发数据、磁场演化模型和最新研究成果，提出"基于磁场剪切梯度与紫外亮度突变的联合预警指标"等创新思路。

3）仿真筛选智能体：高价值假设的科学论证与筛选

该智能体可模拟科学辩论流程，通过多轮迭代优化假设的合理性与新颖性，再采用模拟仿真的方式论证这些假设是否可行，所需资源与时间等，综合评估出最接近当前意图实现的高价值假设，作为后续实测的输入。

4）任务具现智能体：从假设到观测任务的设计

负责将筛选出的假设转化为可执行观测的实验。该智能体基于天文观测约束（如设备可用性、天

气条件、时间窗口）设计最优观测序列与参数，并协调数据采集、预处理与模型计算资源的分配，且能够动态调整方案以应对突发干扰，确保科研任务的高效推进。

5）具身智能望远镜：观测任务的物理执行

作为系统的物理执行层，该智能体直接控制望远镜硬件完成指向跟踪、滤光片切换、曝光参数调整等操作。通过集成传感器数据（如导行误差、大气视宁度），能够实时监测设备状态并实施闭环校正，保障数据采集质量。同时，具备进行多望远镜的网络协同观测的能力，为未来的多智能体（望远镜）组阵协同观测奠定技术基础。

6）元评审智能体：系统级评估审核

通过分析整个假设从生成到仿真到观测到数据的全链条评估审核，评估当前假设、仿真、观测是否符合科学要求，并撰写分析评估报告。若实测验证了当前假设，同时满足了当前意图要求则进入研究总结部分，否则进入反思进化环节。

7）反思进化智能体：反思未达预期因素并迭代优化

模拟同行评审机制，对科学假设、仿真筛选、任务具现等进行批判性审查。其评估维度包括：与现有物理定律的一致性、对异常现象的解释力、可验证性及创新程度。该智能体通过对比历史研究中的失败案例与成功范式，识别假设潜在缺陷，并生成改进建议。在建议基础上通过增量学习融入新的观测数据或物理模型，改进升级算法、模型或参数，如金乌大模型的耀斑爆发预测、黑子形态分类、磁场演化预测方法策略的升级等。

8）研究总结智能体：研究报告的撰写与全过程经验总结

回顾已完成的从意图解析到元评审的整个链路，撰写研究背景与意义、科学假设尝试的方法、仿真策略、观测任务的设计与实现，以及如何调整升级等的研究报告，同时总结过程中的成败经验，为下一次研究提供参考。

通过上述模型与智能体的设计，SIDEST 形成了意图解析-科学假设-仿真筛选-任务具现-具身观测-元评审-反思进化/研究总结的工作流，在工作流中这些智能体并非线性工作，而是处在一个动态的、相互协作与竞争的循环中，智能体通过类似“科学讨论”的方式来挑战和完善彼此的观点，最终通过排序和进化过程，筛选出最优的假设进行模拟仿真，之后驱动望远镜实现真实观测验证并持续优化迭代，实现演进式自主研究。

### 3.2.3 自主演进的闭环系统

SIDEST 系统采用三级闭环架构，实现从科学意图到研究进化的全流程自主化。该系统包含一个全系统闭环和两个内部嵌套闭环，共同支撑 SIDEST 的科研闭环演进。全系统闭环覆盖从意图解析到研究总结的完整科研链条，包含了意图解析-科学假设-仿真筛选-任务具现-具身观测-元评审-反思进化/研究总结的全流程，该闭环强调“人在回路”的交互机制，确保科学家可对关键环节进行监督与干预，

该闭环使 SIDEST 具备从已有研究中提炼新科学问题、更新研究范式的能力；内部嵌套闭环一，以科学假设论证闭环为目标，通过仿真环境对假设进行多轮可行性论证，该环节利用计算仿真替代传统人工推导，快速筛选出具备科学价值且可验证的研究假设，提升科学目标的合理性与可实现性，为后续观测任务奠定；内部嵌套闭环二，以高质量观测闭环为目标，在任务具现与具身观测阶段系统通过实时比对观测数据与任务目标，动态调整观测策略与设备参数，提升望远镜在多变观测场景下的自适应能力，为下一代智能科学观测设施提供了可扩展、可复用的架构范式。

SIDEST 系统的核心在于其动态辩论机制、多源知识融合与持续演化能力，能够发挥科学家在研究过程中目标设定与结果价值判定的作用，在保障科学严谨性的同时提升探索效率，形成“科学意图引导+机器高效执行”的 AI 协同科研新范式。

# 3 系统实现与部署

SIDEST 系统的部署与实现是一个融合自动控制、机器人、人工智能技术和科学研究方法的综合工作，其方法核心在于一个多智能体架构，该架构模拟了科学研究的观测规划、分析推理过程，并通过在线强化学习等技术实现自我迭代优化，该系统将复杂的科研任务分解，由多个各司其职的智能体协作完成。系统实现与部署的基本过程：

1）具身智能望远镜构建：具身智能望远镜是实现当前 SIDEST 的基础，为了验证基于具身智能望远镜的构想，我们在子午工程二期全日面矢量磁像仪（SFMM）上设计了具备感知、交互能力和自动化观测执行的总控系统，实现了“有人值守、无人操作”的智能化运行。

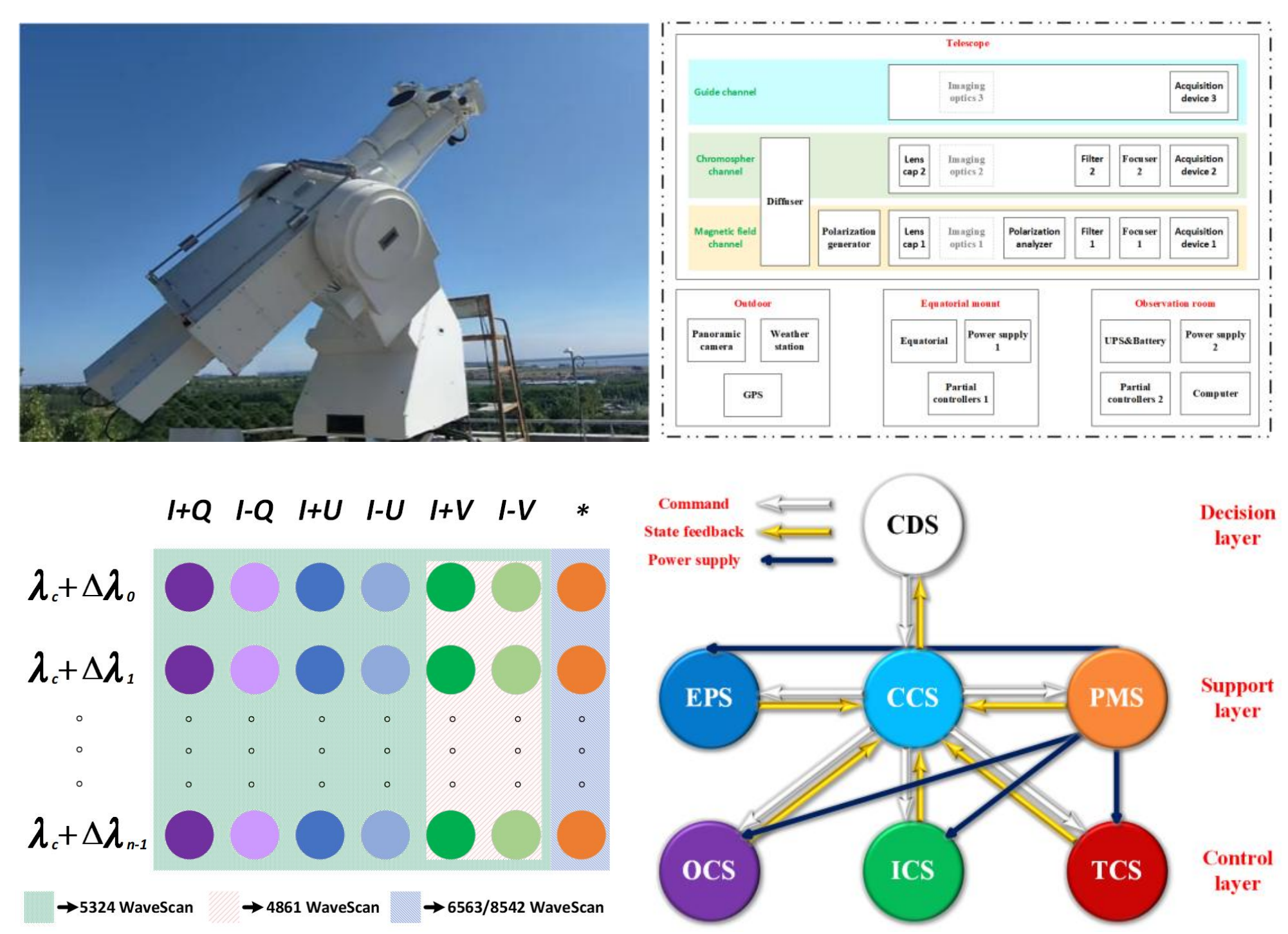

图 2 $a\ SFMM$望远镜　$b\ SFMM$的系统硬件构成
$c\ SFMM$的观测流程　$d\ SFMM$总控架构组成

Fig. 2 $a\ SFMM\ Telescope$　$b\ The\ hardware\ composition\ of\ SFMM$
$c\ Observation\ process\ of\ SFMM$　$d\ SFMM's\ master\ control\ architecture$

在 SFMM 总控系统中设计了环境感知系统（EPS），如全天云量监测与分析系统，能够根据当前全天云量与气象站数据实现环境感知，为系统动作决策提供环境参数支撑；通过望远镜控制系统（TCS）、焦面控制系统（ICS）、电源管理系统（PMS）等系统获得望远镜当下的运行状态、参数，为系统驱动决策提供设备参数支撑；

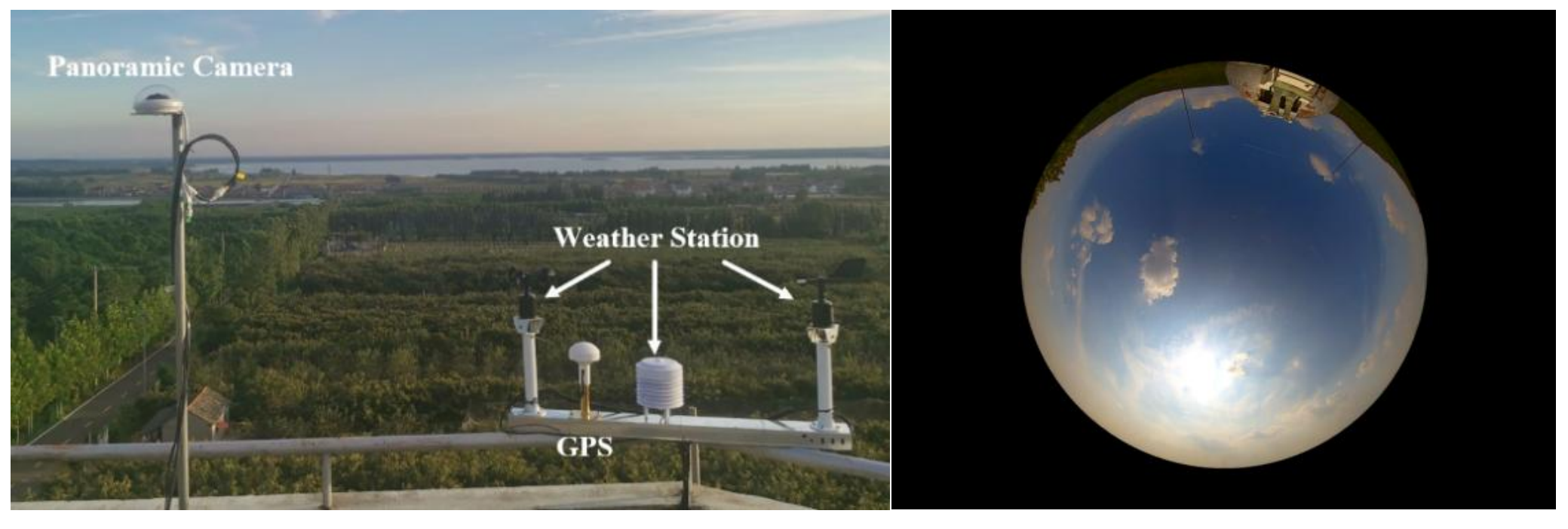

图 3 $a\ SFMM$的气象站与全景相机　$b\ SFMM$全天云量识监测

Fig 3 $a\ SFMM's\ Weather\ Station\ and\ Panoramic\ Camera$　$b\ SFMM's\ Total\ Cloud\ Amount\ Monitor$

基于微服务架构开发了软件系统，通过通信控制子系统（CCS）串联其他子系统到中枢决策子系统(CDS)中，综合观测任务，评估望远镜状态(CDS 右侧为望远镜全部子系统状态)，实现高效的系统参数监测与任务调度。

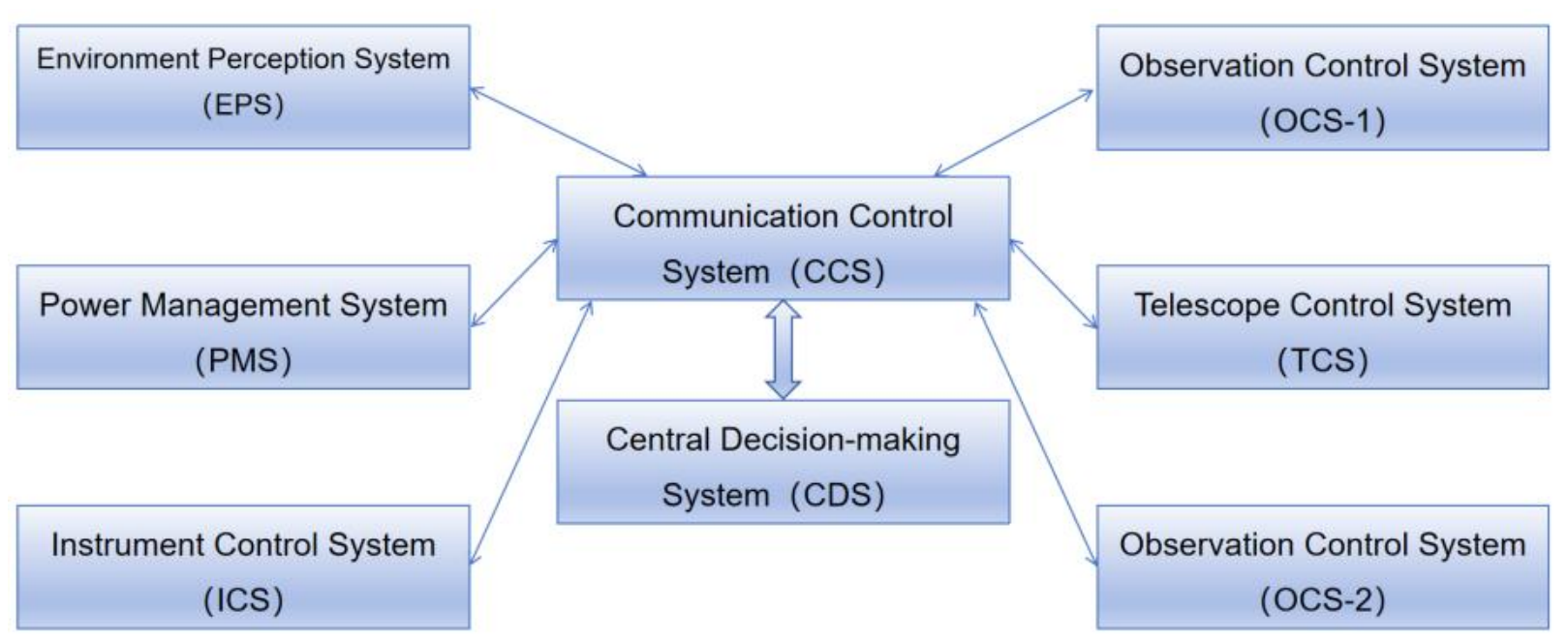

图 4 SFMM 各个子系统通信图

Fig 4 *SFMM's subsystem communication diagram*

目前，该望远镜已具备基于气象环境条件、望远镜运行参数状态感知与观测目标需求等，实现具身观测任务。

2）系统设计与仿真环境构建：构建高可靠、可测试的软硬件基础框架，明确系统的自主能力等级，选定适配的计算硬件、传感器并合理的分配计算资源。设计标准化的通信接口，确保 AI 代理能够与望远镜控制系统（TCS/ICS）无缝交互，避免指令冲突或数据丢包。仿真平台是后续模型训练与验证

的基础，能显著降低直接操作真实设备的风险。

3）模型训练与验证：一方面利用历史科学数据构建面向领域科学问题的基础模型；另一方面利用望远镜系统历史观测日志和专家操作数据，通过模仿学习使 Agent 掌握基础观测流程。进一步引入强化学习，设置以数据质量、观测效率为核心的奖励函数，优化多目标调度与异常响应策略。

4）渐进式部署策略：部署过程需遵循风险可控的渐进原则[24]。首先采用影子模式，使 AI 系统并行接收望远镜数据并生成决策建议，但仅供人类操作员参考，不直接控制硬件，以此评估决策准确性。随后进入人机协同模式，授权 AI 自主执行低风险任务（如平暗场定标等），但在关键操作（如目标切换、异常中断）时需人工确认。最终，在系统通过长期稳定性测试后，可开放完全自主模式，允许其在预设条件下独立完成观测闭环，人类角色转为目标设定与结果监督。这一过程需建立明确的回滚机制，以应对突发故障。

5）建立实时闭环与持续迭代的演进系统，系统正式运行后，将观测图像、传感器数据流持续输入 SIDEST 系统模型，实现系统模型、智能体参数的动态调整（如根据视宁度自适应曝光参数）。基于在线学习框架，利用新观测数据、新的研究文献等持续优化模型，并形成新的研究报告，总结研究结果、形成科学发现，定期评估系统表现，结合人类专家的反馈对决策逻辑进行迭代升级，最终形成“观测-学习-优化”的自主演化循环。

为了实现低延迟决策执行，部分模型应在望远镜侧部署，且需要对模型进行压缩、剪枝或蒸馏，以适应边缘设备有限的计算和存储资源。同时，为了实现实时推理，确保在观测环境或者观测目标快速演化时能够及时作出反应，需要配置高效的计算硬件（如 GPU、FPGA 等）和优化的软件框架。

# 4 实验设计与结果

在 SIDEST 的设计中，主要研究内容包括科学意图的理解与假设论证、具身智能观测以及评估迭代三部分，其中，科学意图理解与假设生成部分，谷歌等团队已验证了一系列基于大模型的可行方法；具身观测部分，我们在 SFMM 上初步实现了一套可行的基础方案；目前，SIDEST 系统中仍需论证 LLM 基于假设的仿真以及基于结果的迭代优化能力。为了验证这两部分的技术路线，同时避免复杂科学难题造成的假设空间爆炸以及无法仿真的问题，我们选择较成熟的工程领域来验证该设计思路，即在精密温控领域设计一个 AI 工程师的进行试验性探索。精密温度控制是科学仪器（太阳望远镜滤光器精密温控）、高端制造（如半导体光刻）、航空航天及生物工程等领域的重要基础保证。这类系统通常表现为高度非线性、大惯性与纯滞后、多变量、强耦合以及易受内外扰动影响的复杂动态特性。传统的精密温控方法，如 PID（比例-积分-微分）控制及其变体（如模糊自适应 PID、增量式 PID）虽然在特定工况下表现良好，但是这类系统的实现也常常面临着一些瓶颈，例如对于不同工况需要展开针对性的研究和调试、学习过程耗时、模型依赖性强、适应性与泛化能力有限等问题。

为了实现基于大模型的自主化分析和智能控制的精密温控系统，我们提出了三种由浅入深、风险可控的实现方案：1）大模型作为整定器或者优化顾问：在离线或非实时场景下，利用大模型强大的知识整合与推理能力，为传统控制器（如 PID）提供参数优化建议；2）大模型作为高级监督与决策层：构建分层控制架构，由大模型负责高层级的策略制定、异常诊断、目标设定以及控制策略代码的实现，底层则仍由封装的 MPC 控制器执行实时控制。3）端到端的大模型控制器：这是一种更为前瞻性的方案，将大模型直接作为控制器（替代常规的单片机、计算机等上位机），输入传感器数据，输出执行器指令，实现对复杂、非线性系统的自适应控制。

第一种方案无法跳出 PID 算法本身的局限性（例如部分适合模糊控制的工况）。第三种方案由于大模型的离散、概率输出等特性，能够实现长期稳定的精密控制目前仍有争议。因此在实验中，我们采用第二种方案——基于 SIDEST 中科学意图理解与假设论证的构想设计的部分智能体，集成 LangGraph 工作流引擎、RAG 检索引擎、MCP 协议管理器、独立温度监控等技术，实现了一个 AI 工程师系统，其架构如图 2 所示。

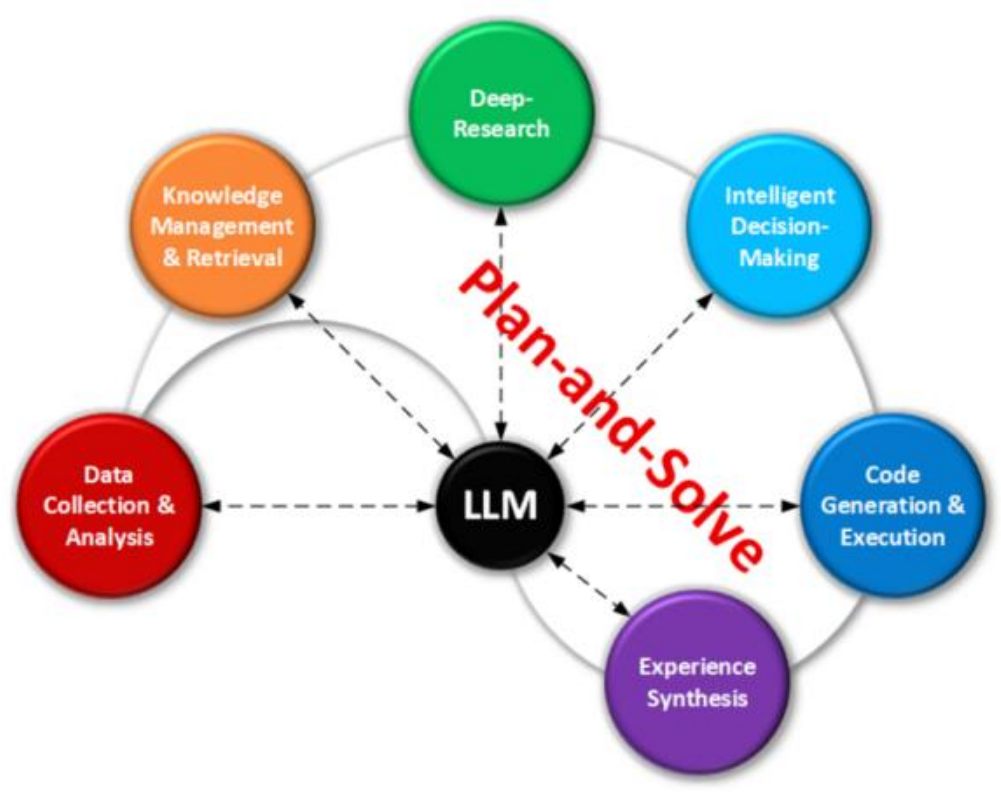

图5 AI工程师系统架构
Fig. 5 AI Engineer System Architecture Diagram

## 4.1 实验设计

在怀柔太阳观测基地全日面太阳磁场望远镜现有滤光器精密温控系统的基础上，为满足特定实验需求，本研究对其进行了适应性改造。核心修改在于：关闭了下位机系统中的闭环精密温控算法（如积分分离 PI 控制），仅保留了高精度温度采集（如使用 24 位 AD）、加热控温驱动以及必要的硬件保护机制。同时，为确保系统安全，在软件层面针对大模型输出的控制量设置了限位保护，系统架构如图 3 所示。上位机中集成了基于大模型的分层控制架构，通过设计智能体工作流，实现采集温度数据并分析其特征、文献查阅、深度研究，并在上位机中撰写控温代码，通过驱动接口进行闭环控制，最终实现滤光器精密温控功能。

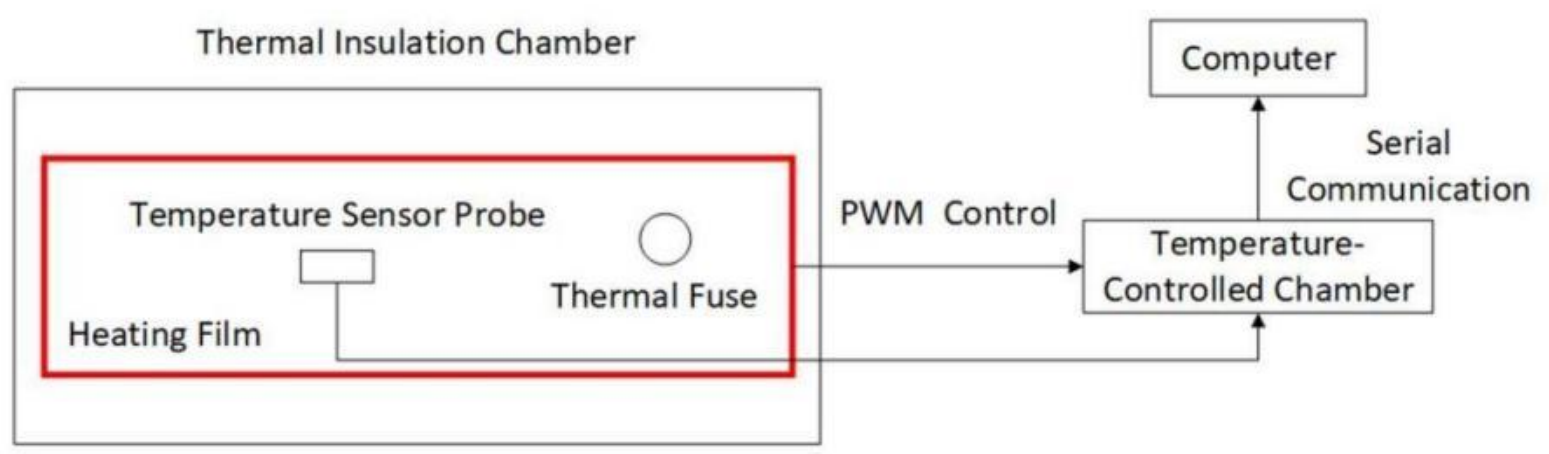

图6 精密温控系统示意图

Fig. 6 Precision Temperature Control System Diagram

该实验系统通过大语言模型、智能体工作流等组合实现，具体过程如下：

第一步，利用系统认知智能体通过硬件接口采集、控制并分析数据特点和变化趋势。

第二步，利用基础知识管理智能体系统分析本地知识库，形成系统基础的控制认知与策略。

第三步，利用深度研究智能体通过本地知识库并结合网络开展深度研究，为系统规划最新的控制策略。

第四步，利用控制策略智能体：基于策略模板、MCP 拓展及深度研究的新策略等，形成当前最佳控制策略。

第五步，利用代码生成与仿真检验智能体：基于最佳控制策略，深度调研并参考近似课题的开源库资料，在沙盒环境生成面向本课题研究的代码，经安全检验后执行代码测试。

第六步，研究总结智能体基于本次研究，开展经验总结、奖励评分，并根据控制稳定度结果是否达到预期的水平决定进一步的研究策略。如果已经达到则保持当前策略，更新策略库。如果尚未到达预期目标，则重复第三步~第六步，实现系统能力的自我演进。

## 4.2 实验结果

基于 4.1 的 AI 温控工程师设计，我们对太阳望远镜滤光器进行了恒温控制测试，目标温度 32.7°C，实验室环境温度约 25°C，最终控温结果显示控制系统的稳定精度峰峰值为±0.0015°C以内（如图 4 所示），达到了滤光器精密温控系统的要求。这一结果说明基于 SIDEST 设计思路的 AI 温控工程师能够通过自主研究-调研文献、提出设计恒温控制算法和软件，并通过反复迭代实现精密的恒温控制系统这一研究目标。

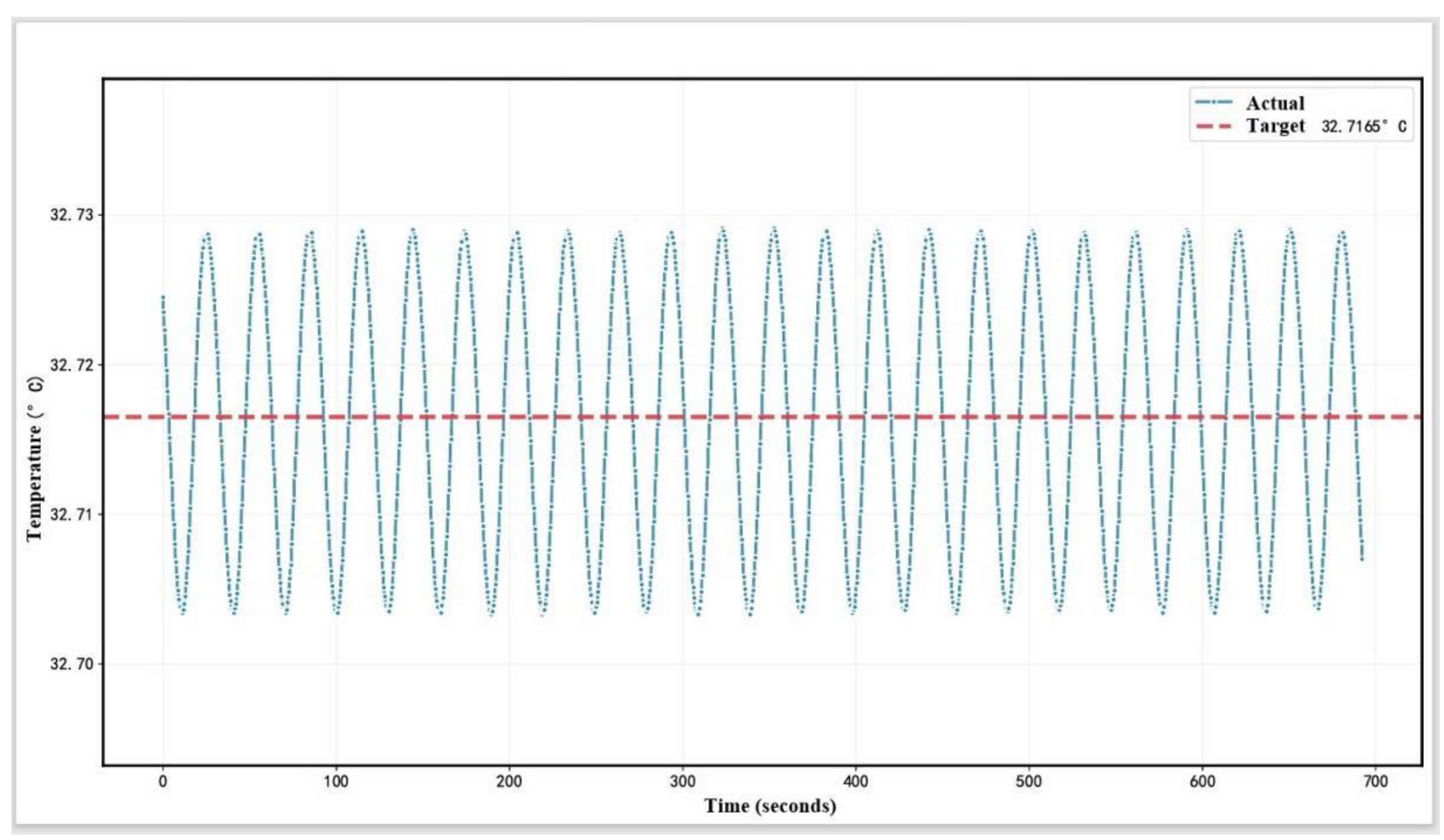

图7 精密温控结果

Fig. 7 Results of Precision Temperature Control

本系统从开始设计到实现精密温控仅耗时三周，而培养一名能够开发精密温控系统软件的工程师通常需要数年时间，充分体现了 SIDEST 框架在实现 AI 工程师方面的高效率，这也让我们看到该框架在科学研究中可能的潜力。在系统设计中，我们采用模块化架构，将不同传感器与驱动器硬件封装为统一的 MCP 服务模块，从而可在同一框架下快速拓展至其他类型的控制系统，例如运动控制平台。因此，该 AI 工程师系统具备良好的可扩展性与推广潜力，其原型模块为 SIDEST 系统的实现提供了可靠的技术基础。

# 5 总结与展望

科学意图驱动的具身智能太阳望远镜系统（SIDEST）提出了一种面向自主科研的新一代望远镜系统架构。本文系统阐述了其设计理念与组成结构，并通过构建太阳望远镜滤光器精密温控的原型系统验证了技术路线的可行性。相较于现有自动化科研平台，SIDEST 的创新主要有以下两个方面：

1）面向科学意图驱动的软硬件协同架构：系统以怀柔观测基地的三台太阳望远镜为硬件基础，构建了一个可响应广域科学意图的望远镜组合网络。在软件层面，设计了三级嵌套闭环的智能体架构，实现从科学意图解析与论证、观测任务具现到系统自我演化的全流程自主科研能力。SIDEST 架构设计支持多种设备的智能化组网协同，为构建 AI 驱动的、网络化的超级天文台（AI 赋能的高端科学仪器[33]、超级实验室）探索技术路线。

2）人类与 AI 深度协同的科研新范式：该系统确立了人类科学家与 AI 科学家深度互补协作的新型研究模式。在此范式下，人类科学家负责提出关键科学问题并研判研究成果的价值，而 AI 科学家系统则承担观测执行、数据分析和模型优化等高重复性、高强度任务，二者形成高效协同的闭环。SIDEST 的系统设计中，从意图输入到结果判定以及中间的过程均设计保证“人在回路”的交互接口，确保人类科学家在可以过程中发挥目标设定与价值判定的关键作用。这种范式既能保持科学研究中人

类在创造性思维和价值判断方面的主导地位，又能够发挥 AI 系统在数据处理和模式识别方面的优势，为更加高效的科学研究、发现提供可能。

尽管该概念从理论构想迈向工程实践仍面临多重挑战，例如在系统集成与性能优化方面，具身智能系统对望远镜硬件的稳定性、控制精度及响应速度提出了极为严苛的要求。复杂的天文观测环境需建立高可靠性的软硬件协同控制机制，以保障观测数据质量与系统鲁棒性。在模型与知识层面，领域大模型的构建依赖于大规模高质量标注数据，然而当前太阳物理领域的标注资源仍相对匮乏。此外，如何将领域知识有效嵌入模型，并确保其推理过程具备逻辑一致性、物理合理性、可解释性，同时减少幻觉现象，仍是需要深入探索的关键问题；同时，如何对自动化科研中的知识产权归属进行准确划分也是需要关注的一个重要科研伦理问题。我们相信随着人工智能技术的快速发展，上述问题会得到更广泛的关注和持续的攻关，相关技术难题也会被逐步识别并系统性地解决。

AI 正在深刻变革着科学研究的范式。我们提出 SIDEST 概念就是希望从技术发展趋势出发用 AI 重新定义天文望远镜、探索天文研究的新范式——未来的望远镜（或者科学仪器）将不再仅仅是被动执行指令的科研工具，而是能够理解科学目标、主动开展探索并具备持续学习能力的科研伙伴。在 AI 的辅助下，科学发现的范式也将逐步从依赖偶然灵感与直觉，向系统化、可规划的“工程化”模式演进。其核心在于，通过大规模历史数据分析与在线实验探索，系统性地识别知识空白、挖掘潜在模式组合，并主动定位高价值的突破方向。这一转变旨在构建一个更加高效、可重复且能持续拓展人类认知边界的科研新范式[25]。在这种 AI 驱动的新范式下，科学研究的参与门槛将大幅降低。科学研究将不再局限于接受过数十年专业训练的专业科学家群体，任何具备科学兴趣与探索精神的个体，都有可能通过自然语言交互的方式，直接参与乃至主导研究进程[26]。这一转变将从根本上拓宽科学研究参与者的范围，并推动科研范式向更加开放、更加协作的方向发生深刻变革。

# 参考文献

[1] Pesnell, W. D., Thompson, B. J., & Chamberlin, P. 2012, The solar dynamics observatory (SDO) (Springer)

[2] Gan, W.-Q., Zhu, C., Deng, Y.-Y., et al. 2019, Advanced Space-based Solar Observatory (ASO-S): an overview, Research in Astronomy and Astrophysics, 19(11), 156

[3] Huang, K., Hu, T., Cai, J., Pan, X., Hou, Y., Xu, L., Wang, H., Zhang, Y., & Cui, X. (2023). Intelligence of Astronomical Optical Telescopes: Present Status and Future Perspectives. Journal of Astronomical Instrumentation, 12(4), 1-45. https://doi.org/10.1234/example-doi

[4] Tong, L. Y., Sun, Y. Z., Yang, X., et al. 2024. Design and application of an autonomous Master ControlSystem for a multi-layer magnetic and helioseismic telescope. Astronomical Techniques and Instruments, 1(3): 1－10.https://doi.org/10.61977/ati2024020.

[5] Bloom N, Jones C I, Van Reenen J, et al. Are ideas getting harder to find?[J]. American Economic Review, 2020, 110(4): 1104−1144.

[6] Frank M C. Baby steps in evaluating the capacities of large language models[J]. Nature Reviews Psychology, 2023, 2(8): 451−452.

[7] Brown T, Mann B, Ryder N, et al. Language models are few-shot learners[J]. Advances in Neural Information Processing Systems, 2020, 33: 1877−1901.

[8] Jiang L Y, Liu X C, Nejatian N P, et al. Health system-scale language models are all-purpose prediction engines[J]. Nature, 2023, 619(7969): 357−362.

[9] Singhal K, Azizi S, Tu T, et al. Large language models encode clinical knowledge[J]. Nature, 2023, 620(7972): 172−180.

[10] Mirza A, Alampara N, Kunchapu S, et al. Are large language models superhuman chemists?[J/OL]. arXiv preprint arXiv:2404.01475, 2024[2025-10-18]. https://arxiv.org/abs/2404.01475 .
[11] Kramer S, Cerrato M, et al. Automated scientific discovery: from equation discovery to autonomous discovery systems[J/OL]. arXiv preprint arXiv:2403.12345, 2024[2025-10-18]. https://arxiv.org/abs/2403.12345 .
[12] Silva R G L. The advancement of artificial intelligence in biomedical research and health innovation: challenges and opportunities in emerging economies[J]. Globalization and Health, 2023, 19(1): 45.
[13] Gottweis J, Weng W H, Daryin A, et al. Towards an AI co-scientist[J/OL]. arXiv preprint arXiv:2502.18864, 2025[2025-10-18]. https://arxiv.org/abs/2502.18864 .
[14] Wang L, Ma C, Feng X, et al. A survey on large language model based autonomous agents[J]. Frontiers of Computer Science, 2024, 18(6): 186345.
[15] Kulkarni A, Alotaibi F, et al. Scientific hypothesis generation and validation: methods, datasets, and future directions[J/OL]. arXiv preprint arXiv:2310.12345, 2023[2025-10-18]. https://arxiv.org/abs/2310.12345 .
[16] Ai, G.-x., & Hu, Y.-f. 1986, The Propose for the Solar Magnetic Field Telescope and Its Working Theorem, Acta Astronomica Sinica, (2), 91−98
[17] Zhang, H. Q., Wang, D. G., Deng, Y. Y., et al. 2007, Solar Magnetism and the Activity Telescope at HSOS, , 7(2), 281−288
[18] S. Djorgovski. "The Roles of Small Telescopes in a Virtual Observatory Environment." Astrophysics and space science library
[19] Li Y Y , Bai Y , Wang C ,et al.Deep Learning and LLM-based Methods Applied to Stellar Lightcurve Classification[J]. 2024.DOI:10.34133/icomputing.0110.
[20] Bu K, Liu Y C, Ju X L. Efficient utilization of pre-trained models: A review of sentiment analysis via prompt learning[J]. Knowledge-Based Systems, 2024, 283: 111148.
[21] Elkhatib Y, Coulson G, Tyson G. Charting an intent driven network. In:Proc. of the 201713th Int'l Conf. on Network and Service Management (CNSM). IEEE Computer Society, 2017.
[22] King R D, Whelan K E, Jones F M, et al. Functional genomic hypothesis generation and experimentation by a robot scientist[J]. Nature, 2004, 427(7059): 247−252.
[23] 林佳本, 沈洋斌, 朱晓明等. 怀柔太阳观测基地全日面磁场自动化观测系统[J]. 天文研究与技术, 2013, 10(4): 392-396.
Lin Jiaben Shen Yangbin Zhu Xiaoming et.al, Design of the Automatic Observing System for Full Disk Magnetogram in HSOS,Astronomical Research and Technique, 2013, 10(4): 392-396.
[24] Cunshi Wang, Xinjie Hu, Yu Zhang, et.al. StarWhisper Telescope: Agent-Based Observation Assistant System to Approach AI Astrophysicist.(2024-11). https://arxiv.org/abs/2412.06412
[25] 刘志毅， 第五范式， 中信出版社， 2025
Liu Zhiyi, The Fifth Paradigm, Citic Press Corp.,2025
[26] 毛永娜,向娥,李瑀旸,王存是,刘继峰.人工智能驱动的天文科学教育体系构建与实践探索.中小学科学教育, 2025,2(2):41-46
Mao Yongna, Xiang E, Li Yuyang, Wang Cunshi, Liu Jifeng, Artificial Intelligence-Driven Construction and Practical Exploration of an Astronomical Science Education System, Science Education for Primary and Secondary Schools2025,2(2):41-46
[27] MITCHENER L, YIU A, CHANG B, et al. Kosmos: An AI scientist for autonomous discovery. arXiv, 2025. https://arxiv.org/pdf/2511.02824
[28] Swanson K, Wu W, Bulaong NL, Pak JE, Zou J. The Virtual Lab of AI agents designs new SARS-CoV-2 nanobodies. Nature. 2025 Oct;646(8085):716-723. doi: 10.1038/s41586-025-09442-9. Epub 2025 Jul 29. PMID: 40730228.
[29] Tang, J., Xia, L., Li, Z., & Huang, C. (2025). AI-Researcher: Autonomous Scientific Innovation. arXiv. 2025 https://arxiv.org/abs/2505.18705
[30] http://www.aiar.xjtu.edu.cn/info/1004/3785.htm
[31] Shao, M., Liu, S., Xu, H., Jia, P., Wang, H., Tong, L., Bai, Y., Yang, C., Li, Y., Li, N., & Lin, J. (2025). Advances and Challenges in Solar Flare Prediction: A Review. https://arxiv.org/pdf/2511.20465
[32] 白春礼. AI赋能高端科学仪器 推动智能科研跃迁发展[J]. 世界科技研究与发展, DOI：10.16507/j.issn.1006-6055.2025.11.001

Chunli Bai,AI Empowers High-end Scientific Instruments to Drive the Leap-forward Development of Intelligent Scientific Research[J]. World Sci-Tech R & D. DOI：10.16507/j.issn.1006-6055.2025.11.001